\documentclass[useAMS,usenatbib]{mn2e}
\usepackage{myaasmacros}
\usepackage{amssymb}
\usepackage{graphicx}
\usepackage{hyperref}
\usepackage{xcolor}
\usepackage{amsmath}
\usepackage{amsfonts} 
\usepackage{bm}

\def\rhodm{\rho_\mathrm{dm}}
\def\rhotot{\rho_\mathrm{tot}}

\def\rhos{\rho_\mathrm{s}}

\def\vztwo{\sigma_{z}^2}

\def\vztwoi{\sigma_{z}^2}
\def\vztwon{\sigma_{z,n}^2}
\def\vzstwoj{\sigma_{z,j}^2}
\def\vzstwojn{\sigma_{z,j,n}^2}

\def\rhoeff{\rho_\mathrm{dm}^\mathrm{eff}}
\def\beq{\begin{equation}}
\def\eeq{\end{equation}}
\def\msun{\,M$_\sun$}
\def\bit{\begin{itemize}}
\def\eit{\end{itemize}}
\def\ben{\begin{enumerate}}
\def\een{\end{enumerate}}

\title[The local dark matter density from K dwarfs]
{A new determination of the local dark matter density from the kinematics of K dwarfs}

\author[S. Garbari et al.]{Silvia Garbari$^{1}$\thanks{e-mail:
\texttt{silvia@physik.uzh.ch}}, Chao Liu$^{2,3}$, Justin I. Read$^{4,5}$, George Lake$^1$\\
  $^1$Institute for Theoretical Physics, University of Z\"urich, Winterthurerstrasse 190, 8057 Z\"urich, Switzerland\\
  $^2$Key Lab of Optical Astronomy, National Astronomical Observatories, CAS, 20A Datun Road, Chaoyang District, 100012 Beijing, China\\
  $^3$Max-Planck-Institut f\"ur Astronomie, K\"onigstuhl 17, D-69117 Heidelberg, Germany\\ 
  $^4$Department of Physics and Astronomy, University of Leicester, University Road, LE1 7RH Leicester, UK\\
  $^5$Institute for Astronomy, Department of Physics, ETH Z\"urich, Wolfgang-Pauli-Strasse 27, CH-8093 Z\"urich, Switzerland}

\begin{document}

\pagerange{\pageref{firstpage}--\pageref{lastpage}} 
\pubyear{2012}

\maketitle

\label{firstpage}

\begin{abstract}
We apply a new method to determine the local disc matter and dark halo matter density to kinematic and position data for $\sim 2000$ K dwarf stars taken from the literature. Our method assumes only that the disc is locally in dynamical equilibrium, and that the `tilt' term in the Jeans equations is small up to $\sim 1$\,kpc above the plane. We present a new calculation of the photometric distances to the K dwarf stars, and use a Monte Carlo Markov Chain to marginalise over uncertainties in both the baryonic mass distribution, and the velocity and distance errors for each individual star. We perform a series of tests to demonstrate that our results are insensitive to plausible systematic errors in our distance calibration, and we show that our method recovers the correct answer from a dynamically evolved N-body simulation of the Milky Way. We find a local dark matter density of $\rhodm=0.025^{+0.014}_{-0.013}$\msun pc$^{-3}$ ($0.95^{+0.53}_{-0.49}$\,GeV\,cm$^{-3}$) at 90\% confidence assuming no correction for the non-flatness of the local rotation curve, and $\rhodm=0.022^{+0.015}_{-0.013}$\msun pc$^{-3}$  ($0.85^{+0.57}_{-0.50}$\,GeV\,cm$^{-3}$) if the correction is included. Our 90\% lower bound on $\rhodm$ is larger than the canonical value typically assumed in the literature, and is at mild tension with extrapolations from the rotation curve that assume a spherical halo. Our result can be explained by a larger normalisation for the local Milky Way rotation curve, an oblate dark matter halo, a local disc of dark matter, or some combination of these.

\end{abstract}



\section{Introduction}\label{sec:intro}
The local dark matter density is an average over a small volume, typically a few hundred parsecs, around the Sun. It provides constraints on the local halo shape and allows us to predict the flux of dark matter particles in laboratory detectors. The latter is required to extract information about the nature of a dark matter particle from such experiments, at least in the limit of a few tens to hundreds of detections \citep{Peter_2011}. The Galactic halo shape can be constrained by combining two methods of determining the local dark matter density. Firstly, one can infer it from the Galactic rotation curve ($\rho^\mathrm{ext}_\mathrm{dm}$). This requires an assumption about the shape of the Galactic halo \citep[typically spherical; e. g.][] {sofue_unified_2008,weber_determination_2010,catena_2010}. Secondly, one can calculate the dark matter density locally from the vertical kinematics of stars near the Sun ($\rho_\mathrm{dm}$) \citep[e. g.][]{bahcall_self-consistent_1984,holmberg_local_2000}. If $\rho_\mathrm{dm}<\rho^\mathrm{ext}_\mathrm{dm}$, this suggests a prolate dark matter halo for the Milky Way; while  $\rho_\mathrm{dm}>\rho^\mathrm{ext}_\mathrm{dm}$, could imply either an oblate halo or a dark disc \citep{lake_must_1989,read_thin_2008,read_dark_2009}. 

Determining the local matter density from the kinematics of stars in the Solar Neighbourhood has a long history dating back to \cite{oort_force_1932,oort_note_1960} in the 1930's. Oort used the classical method of solving the combined Poisson-Boltzmann equations for a sample of stars, assumed to be stationary in the total matter distribution of the disc. He found 50\% more mass than the sum of known components. A more modern study by \cite{bahcall_self-consistent_1984} introduced a new method that described the visible matter as a sum of isothermal components. He also found dynamically significant dark matter in the disc\footnote{We should be careful about what we mean by `dark matter in the disc'. Early studies like \cite{oort_force_1932} were typically interested in missing disc-like matter (a `thin dark disc'); more modern studies try to constrain a significantly more extended dark matter halo that has a near-constant dark matter density up to $\sim 1$\,kpc. Even the `dark disc' predicted by recent cosmological simulations \citep{read_thin_2008,read_dark_2009} is sufficiently hot that its dark matter distribution is approximately constant up to $\sim 1$\,kpc. Throughout this paper when we talk about `dark matter in the disc' we refer to a constant density dark matter component within the disc volume.} \citep{bahcall_k_1984}. Using faint K dwarfs at the South Galactic Pole, \cite{bahcall_local_1992} confirmed his earlier result that more than $50\%$ of the mass was dark, although with a lower statistical significance. However, the early studies by \cite{oort_force_1932,oort_note_1960} and \cite{bahcall_self-consistent_1984,bahcall_k_1984} assumed that different tracers could be simply averaged to form a single tracer population. \cite{kg_1989c} demonstrated that the two samples of F stars analysed by \cite{bahcall_self-consistent_1984} were not compatible (i.e. they had different spatial density distribution, but no evidence for a difference in their kinematics) and therefore should not be averaged. They re-analysed the K giant sample used by \cite{bahcall_k_1984}, assigning more realistic errors to the density profile and using a more detailed fit to the velocity data, finding a value of total matter density compatible with the observed one. They concluded that the determination of the local volume density remained limited by systematic and random errors with the available data.

With the launch of the ESA satellite Hipparcos (1997), the kinematics and position of tracer stars were measured with much higher accuracy. The improved distance measures give a much more accurate measurement of the local luminosity function, so that the total amount of visible matter can be better estimated as well. The latest dynamical measurements of the local density of matter -- $\rhotot$ -- from Hipparcos data show no compelling evidence for a significant amount of dark matter in the disc \citep{creze_distribution_1998,holmberg_local_2000}. \cite{holmberg_local_2000} found $\rhotot = 0.102\pm0.01$M$_{\odot}$pc$^{-3}$, with a contribution of about $0.095$M$_{\odot}$pc$^{-3}$ in visible matter, consistent with the \cite{kg_1989b}'s value.

In addition to the local volume density, several authors have calculated the local {\it surface density} of gravitating matter, probing up to larger heights above the disc plane \citep[typically $\sim 1$\,kpc; e.g.][]{kg_1989a, kg_1989b,kg_1991, holmberg_local_2004}. Using faint K dwarfs at the South Galactic Pole, and using a prior from the rotation curve, \cite{kg_1989b, kg_1991} find $\rhodm^\mathrm{KG} = 0.010\pm 0.005$\,\msun\,pc$^{-3}$, consistent with that expected from the rotation curve assuming a spherical Galactic dark matter halo\footnote{Note that this consistency with the rotation curve is somewhat circular since this is input as a prior in their analysis.} \citep[e. g.][] {sofue_unified_2008,weber_determination_2010,catena_2010}. A similar result was found in the post-Hipparcos era by \cite{holmberg_local_2004}. Recently, \cite{monibidin_2012} have estimated the surface density using tracers at heights $1.5 < z < 4$\,kpc above the disc, making a rather stronger claim (incompatible with the earlier results of \cite{kg_1991} and \cite{holmberg_local_2004}) that there is no dark matter near the Sun. However, \cite{bovytrem_2012} demonstrate that this result is erroneous and owes to one of ten assumptions used by \cite{monibidin_2012} being false. Furthermore, \cite{sanders_2012} estimate that the velocity dispersion gradients derived by \cite{monibidin_2012} could be biased by up to a factor of two, which would also significantly alter their determination of $\rhodm$. 

With next generation surveys round the corner \citep[e.g. Gaia;][]{jordan_gaia_2008}, a significant improvement in the number of precision astrometric, photometric and spectroscopic measurements is expected. For this reason, \cite{paper1} (hereafter Paper I) revisited the systematic errors in determining $\rhodm$ from Solar Neighbourhood stars; these will likely soon become the dominant source of error, if they are not already. We were the first to use a high resolution N-body simulation of an isolated Milky Way-like galaxy to generate mock data. We used these mock data to study a popular class of mass modelling methods in the literature that fit an assumed distribution function to a set of stellar tracers \citep{holmberg_local_2000,binney_galactic_2008}. We found that realistic mixing of stars due to the formation of a bar and spiral arms (similar to those observed in the Milky Way) breaks the usual assumption that the distribution function is separable, leading to systematic bias in the recovery of $\rhodm$. We then introduced a new method that avoids this assumption by fitting instead moments of the distribution function (i.e. that solves the Jeans-Poisson equations). Our Minimal Assumption method (or MA method) uses a Monte Carlo Markov Chain technique (hereafter MCMC) to marginalise over remaining model and measurement uncertainties. Given sufficiently good data, we showed that our method recovers the correct local dark matter density even in the face of disc inhomogeneities, non-isothermal tracers and a non-separable distribution function.

In this article, we apply our MA method to real data from the literature. The key advantages of our new method over previous works are that: (i) we use a `minimal' set of assumptions; (ii) we use a MCMC to marginalise over both model and measurement uncertainties; and (iii) we require no prior from the Milky Way rotation curve as has been commonly used in previous works \citep{kg_1989b,kg_1991,holmberg_local_2000,holmberg_local_2004}. This latter means that we can compare our determination to that derived from the rotation curve to constrain the Milky Way halo shape. Our method requires at least one equilibrium stellar tracer population with known density fall off $\nu(z)$ and vertical velocity dispersion $\vztwo(z)$, both as a function of height $z$. The requirements for a suitable sample of stellar tracers are that: (i) they are in dynamical equilibrium with the Galactic potential (i.e. they must be sufficiently dynamically old to have completed many vertical oscillations through the Galactic plane); (ii) they are available in sufficient numbers to give good statistical precision; (iii) they have reliable distances and vertical velocities $v_z$; (iv) the sample completeness needs to be sufficiently well understood in order to measure the density fall off as a function of the distance $z$ from the disc plane; and (v) they extend up to 2-3 times the disc scale height (in order to break a degeneracy between the disc and dark matter densities; see Paper I). While full six dimensional phase space information is now available for a large number of stars (e.g. RAVE \citealt{steinmetz_rave:_2003,steinmetz_radial_2006,zwitter_radial_2008}; and SEGUE \citealt{yanny_segue:_2009}), these surveys are magnitude rather than volume complete, with additional survey selection effects based on colour. This makes it difficult to reliably estimate $\nu(z)$ for a given tracer population. For this reason, we return to the volume complete K-dwarf data from \cite{kg_1989b} for our disc tracers -- the `KG' data. These data consist of a photometric sample of 2016 K dwarf stars, complete in the $z$-range $\sim 0.2-1.5$\,kpc, with a spectroscopic sample of 580 K dwarfs (most of which are included in the photometric catalogue). We use data from Hipparcos and SEGUE \citep{kotoneva_2002,zhang} to perform a new photometric distance measurement for each K dwarf star. We model the local gravitational potential using the baryonic mass distribution of the Galactic disc by \cite{flynn_2006}.

This article is organised as follows. In Section \ref{sec:data}, we present the K dwarf data from \cite{kg_1989b} (hereafter KG989II) and describe our new distance determinations (\ref{sec:newdist}). In Section \ref{sec:Met}, we summarise our MA mass modelling method (\ref{sec:MA}) and, for comparison, the method adopted by KG89II (\ref{sec:KG}). In Section \ref{sec:Nbody}, we test both methods on a mock data set derived from a dynamically evolved N-body simulation. In Section \ref{sec:Res}, we apply our MA method to the KG data and present our results. Finally in Section \ref{sec:discuss+concl}, we summarise and present our conclusions. 

\section{Data}\label{sec:data}
\begin{figure}
\center
\includegraphics[width=0.45\textwidth]{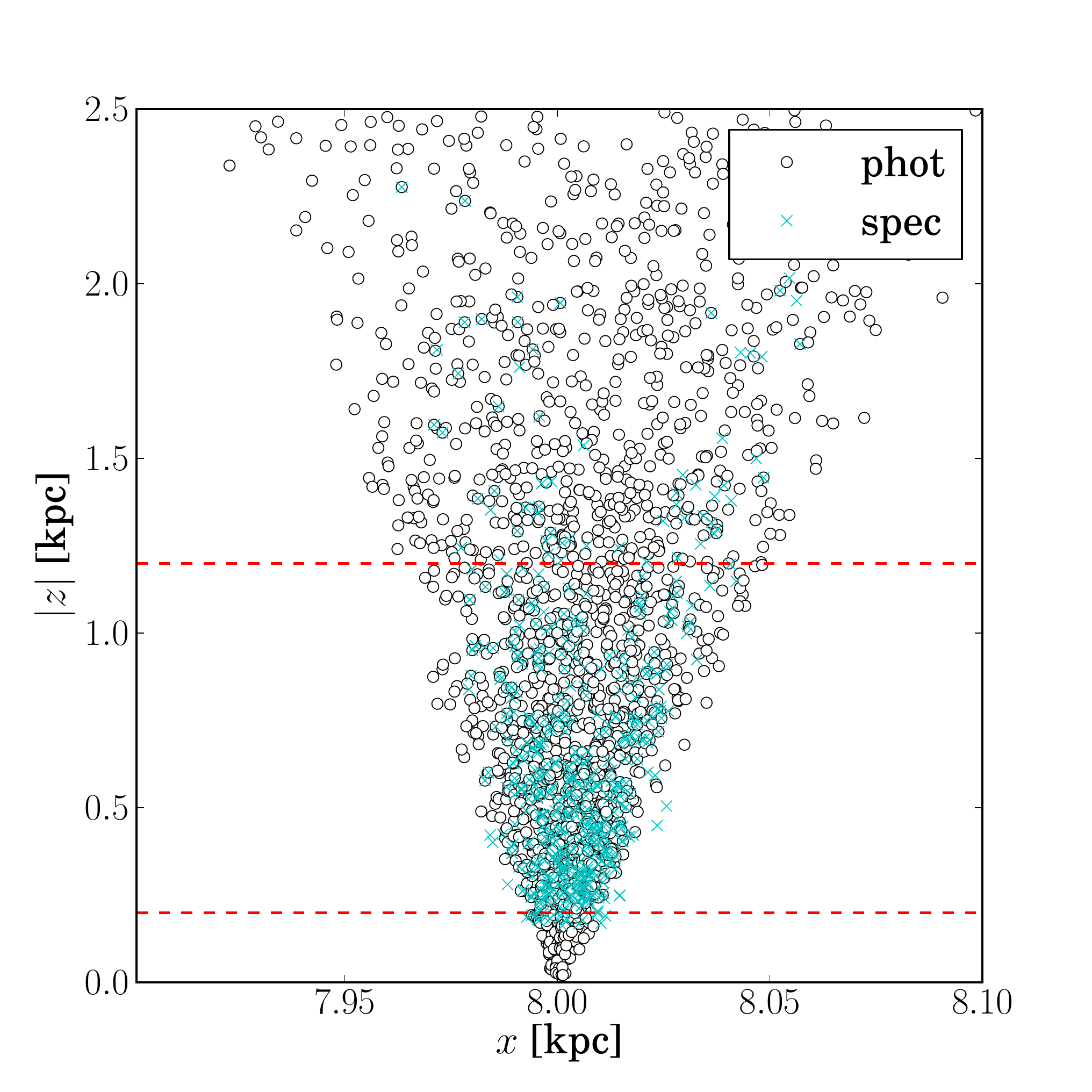}
\caption{Spatial distribution of the K dwarf sample from KG89II. Blue crosses: spectroscopic sample; empty circles: photometric sample. The red dashed lines mark the range of $200<z<1200$\,pc use in our analysis (see Section \ref{sec:newdist}).}
\label{fig:data}
\end{figure}

KG89II present a catalogue of 2016 K stars with photometry in $B$ and $V$ bands, and another of 580 K dwarfs (most of which are also included in the photometric catalogue) including radial velocities, at South Galactic pole (figure \ref{fig:data}). The stellar density fall off of these tracers was derived from star counts. At large $z$, the mean metallicity of the stars is known to decrease below the Solar Neighbourhood value. Such a gradient translates into an absolute magnitude gradient, since the position of the main sequence in the colour-magnitude diagram changes with the metallicity: metal poor dwarfs are fainter than metal rich ones at the same temperature or colour (the opposite is true for giant stars). So, if there is a vertical metallicity gradient, the photometric parallaxes used for the derivation of the density fall off -- $\nu(z)$ -- will be systematically wrong as one moves away from the plane. Unfortunately, the metallicity could not be measured directly for these stars, so KG89II derived the density fall off using an assumed constant metallicity gradient for the K dwarfs. They considered two different gradients to estimate the magnitude of the uncertainties, namely $d\mathrm{[Fe/H]}/dz=0$ (constant metallicity) and $d\mathrm{[Fe/H]}/dz=-0.3$\,dex\,kpc$^{-1}$ (in their analysis KG89II consider this latter as the fiducial metallicity gradient for K dwarfs). 

The tracers' vertical distance determination is fundamental for our analysis. Twenty years after KG89II's study, we can re-calibrate the distances for these stars using modern survey data to estimate the metallicity distribution function of K dwarfs at different $z$, and Hipparcos parallaxes to calibrate the photometric distances. Our distance re-calibration procedure is described, next.

\subsection{A new distance determination for the K dwarfs}\label{sec:newdist}
To calculate the vertical distances $z$ for KG89II's sample, we must derive a relationship between the metallicity $\mathrm{[Fe/H]}$, the vertical distance $z$ and the absolute magnitude $M_V$ of K dwarf stars in the disc. Since the metallicity is not included in KG89II's catalogue, we can only hope to derive a distance distribution function $P_*(z)$ for each star of the sample, based on an observed metallicity distribution function for these stars.  

We consider two different catalogues of K dwarfs with distances and metallicity for our calibration. The first catalogue by \cite{kotoneva_2002} consists of 431 K dwarfs from Hipparcos, representing a complete catalogue of the metal content in nearby K dwarfs extending up to $z\sim 100$\,pc. The vertical distances $z$ for these stars are very accurately determined from Hipparcos parallaxes. The second catalogue by \cite{zhang} contains 5000 SEGUE K dwarfs spanning a much wider range of $z$, namely between $300$ and $2000$\,pc. However, in this case, the distance determination for these stars is much less certain: the distance errors are about 10$\%$.

\begin{figure}
\center
\includegraphics[width=0.45\textwidth]{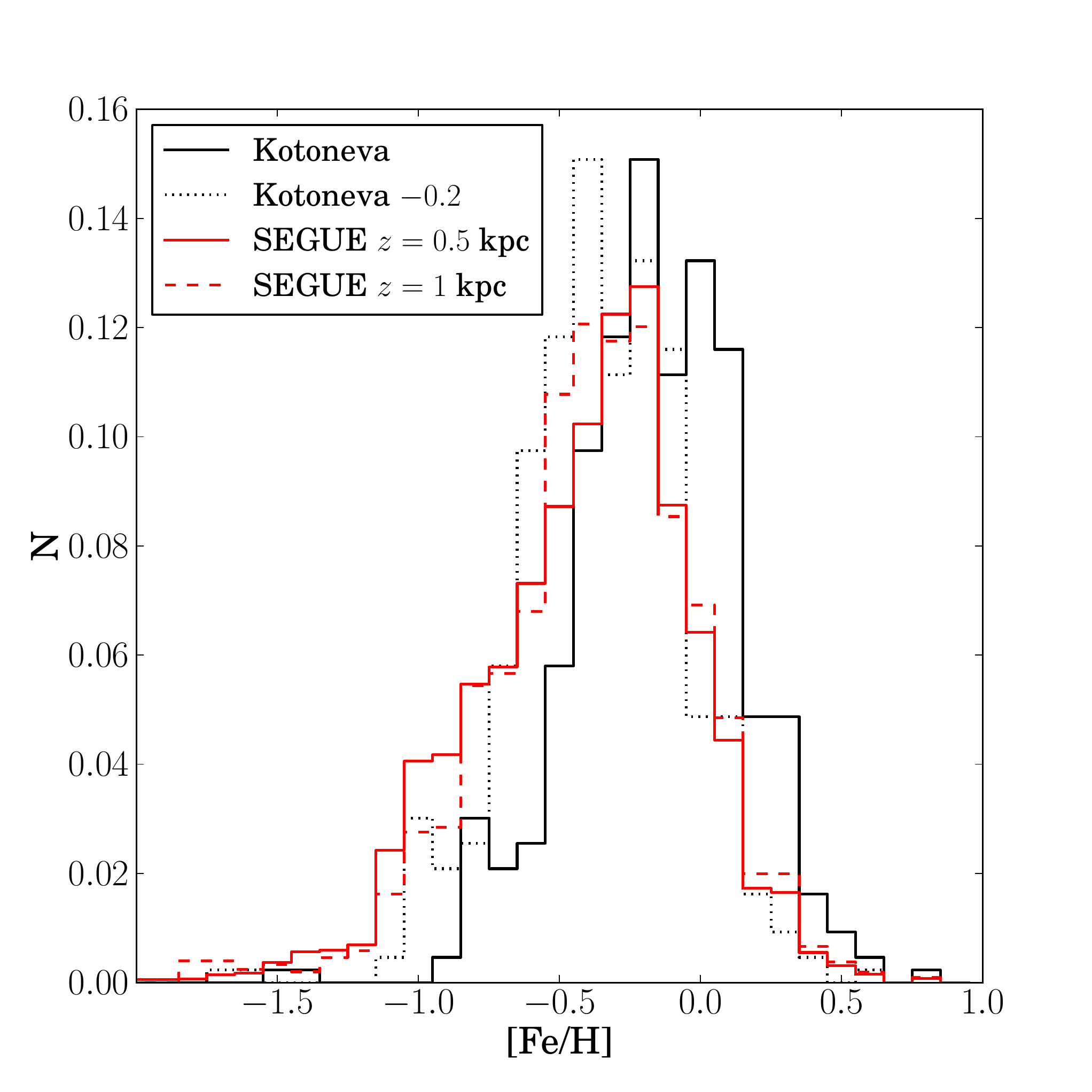}
\caption{The $z \lesssim 100$\,pc MDF (black solid histogram) from \protect\cite{kotoneva_2002}. A shift of 0.2\,dex (black dotted histogram) approximately overlaps the MDF from SEGUE K dwarfs computed at $z \sim 500$\,pc (red solid histogram). The red dashed histogram is the SEGUE MDF at $z \sim 1000$\,pc.}
\label{fig:methist}
\end{figure}

We combine these two catalogues to build the metallicity distribution function (MDF) $Q(\mathrm{[Fe/H]}(z),z)$ for K dwarfs. Comparing the SEGUE metallicity distribution function at $z=500$\,pc with the MDF from \cite{kotoneva_2002} for $z<100$\,pc (see the red and black solid histograms in figure \ref{fig:methist}, respectively), we notice that the two MDFs have very similar shape, but shifted means. There is a vertical metallicity gradient from 0 to 500\,pc of $\sim -0.4$\,dex\,kpc$^{-1}$ (corresponding to a shift of $-0.2$\,dex; see figure \ref{fig:methist} black dotted histogram). At larger height than this, the gradient is weaker: the SEGUE MDF at 1\,kpc (dashed red histogram) is similar to the one at 500\,pc. We adopt the \cite{kotoneva_2002} MDF in the Galactic plane, then we apply a linear metallicity gradient of $\sim -0.4$\,dex\,kpc$^{-1}$ between 100\,pc and 500\,pc to match the SEGUE MDF at $z=500$\,pc; we extend this shifted \cite{kotoneva_2002}'s MDF up to $z=800$\,pc, and we adopt the SEGUE MDF for $z>800$\,pc, as shown in figure \ref{fig:met}. We explore an alternative $Q(\mathrm{[Fe/H]}(z),z)$ for the K dwarfs in Appendix \ref{app:met}. 

\begin{figure}
\center
\includegraphics[width=0.45\textwidth]{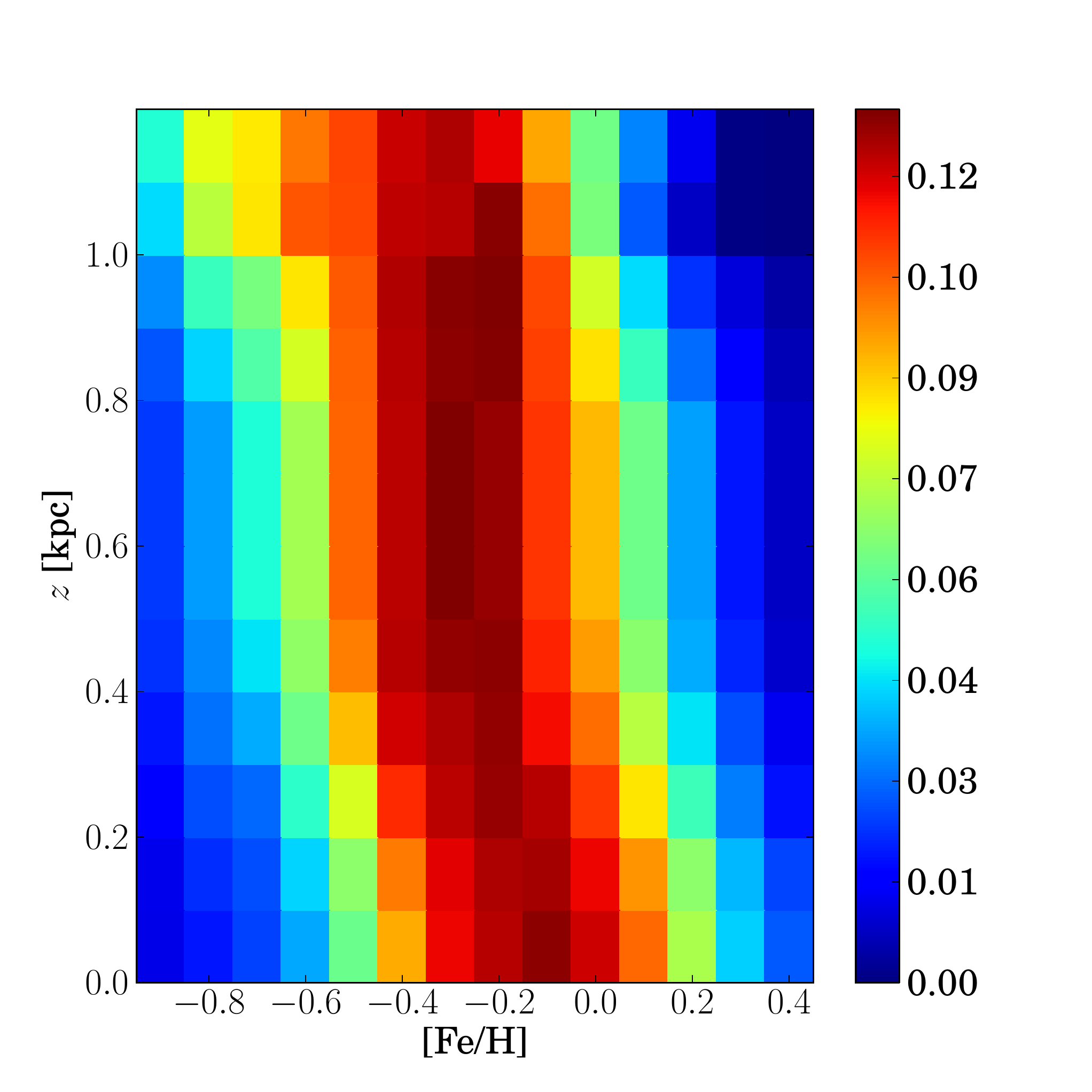}
\caption{Combined MDF of \protect\cite{kotoneva_2002} and SEGUE \protect\citep{zhang}. The colours show the probability values of [Fe/H] given $z$.}
\label{fig:met}
\end{figure}

With the metallicity distribution function $Q(\mathrm{[Fe/H]}(z),z)$ for K dwarfs, we next derive the distance probability distribution $P_*(z)$ for each star of the KG89II's catalogue. This requires calculating the relationship between $z$-distance and metallicity $\mathrm{[Fe/H]}$ for each star.

\begin{figure}
\center
\includegraphics[width=0.45\textwidth]{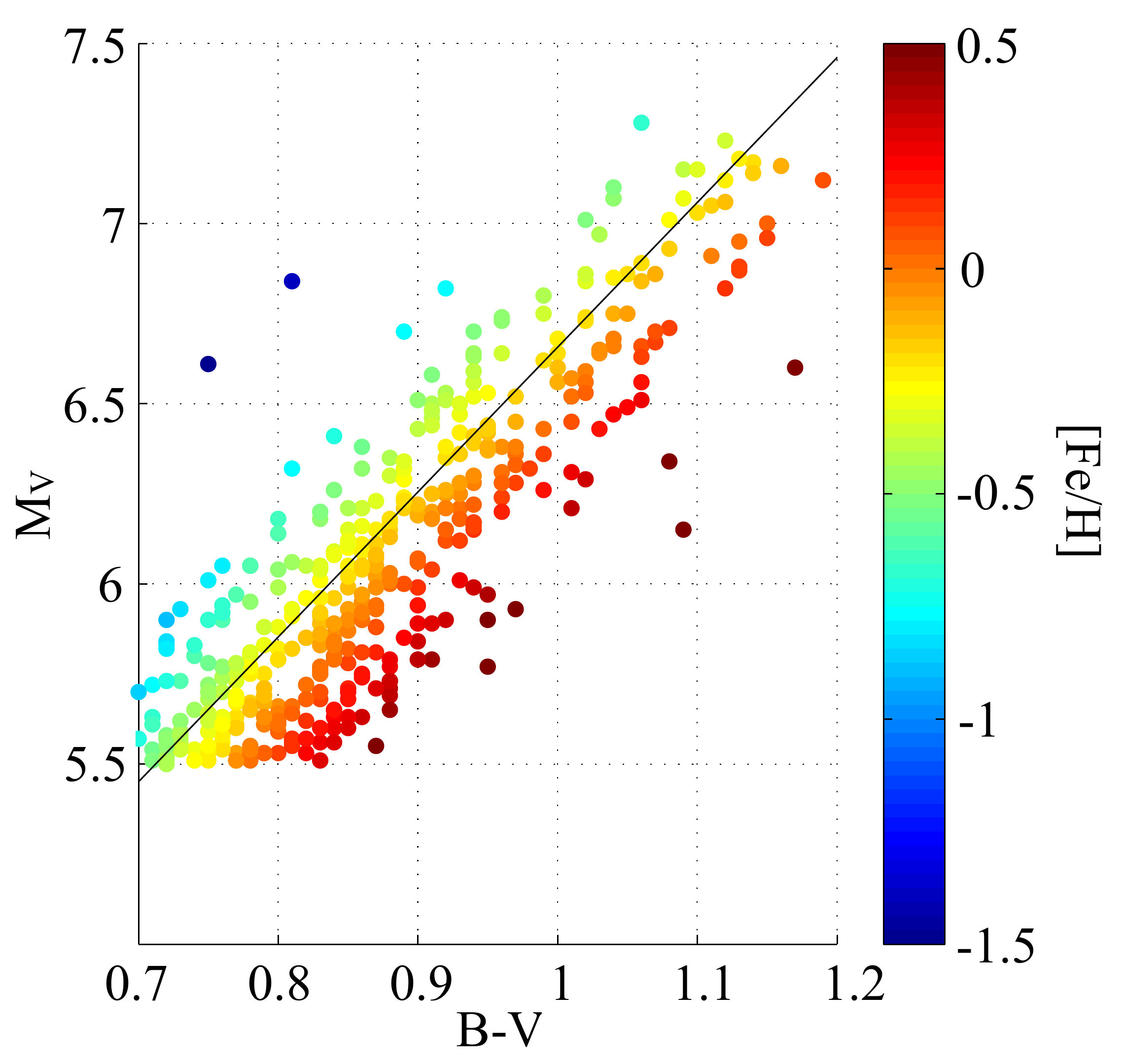}
\caption{$M_V$ as a function of colour $B-V$ and metallicity $\mathrm{[Fe/H]}$ for 431 K dwarfs selected from \protect\cite{kotoneva_2002} (at heights $z < 100$\,pc). The $\mathrm{[Fe/H]}$ is coded by colour. The black line is a linear fit to the data between $\mathrm{[Fe/H]}=-0.7$ and 0\,dex.}
\label{fig:kot}
\end{figure}

Figure \ref{fig:kot} shows the absolute magnitude $M_V$ (ordinate) of \cite{kotoneva_2002}'s K dwarfs as a function of colour index $B-V$ (abscissa) and metallicity $\mathrm{[Fe/H]}$ (colours). We fit this using a polynomial:

\beq
\begin{split}
M_V=& a_{0,0}+a_{1,0}(B-V)+a_{0,1}\mathrm{[Fe/H]}+a_{2,0}(B-V)^2\\
&+a_{1,1}(B-V)\mathrm{[Fe/H]}+a_{0,2}\mathrm{[Fe/H]}^2+a_{0,3}(B-V)^3\\
&+a_{2,1}(B-V)^2\mathrm{[Fe/H]}+a_{1,2}(B-V)\mathrm{[Fe/H]}^2
\end{split}
\label{eq:poly}
\eeq
The best-fit parameters are [$-5.795$, $27.92$, $0.1291$, $-22.74$, $-2.003$, $0.04917$, $7.113$, $1.02$, $-0.04274$], with an error of $\sim 0.03$\,mag. 

Once we have $M_V=M_V(B-V,\mathrm{[Fe/H]})$, we write the distance modulus as:
\beq
d=10^{\frac{V-A_V-M_V+5}{5}}=d(V,B-V,\mathrm{[Fe/H]})\label{eqn:d}
\eeq
where $V$ is the apparent magnitude and $A_V$ is the extinction; we use the value $A_V=0.062$\,mag from \cite{schlegel_1998}, given the mean Galactic coordinates of KG89II's data. The vertical distance $z$ for a single star is then:
\beq
z_*=z_*(l,b,d)=z_*(l,b,B-V,\mathrm{[Fe/H]}) \label{eq:zFe}
\eeq
where $l$ and $b$ are the Galactic longitude and latitude. We know $l$, $b$ and $B-V$ for each star of KG89II's sample, so the only free parameter is $\mathrm{[Fe/H]}$. This means that the vertical distance for each star will be given by a probability distribution $P_*(z)$ corresponding to a metallicity distribution $P_*(\mathrm{[Fe/H]})$ for that star:
\beq
P_*(z)=P_*(\mathrm{[Fe/H]}(z)) \label{eq:pzFe}
\eeq 
In practice, this equation must be solved iteratively because $\mathrm{[Fe/H]}$ is itself a function of $z$ through equation \ref{eq:zFe}. The iterative process proceeds as follows: 
\ben
\item We start the first iteration by assuming that the distance distribution function of a single star is $P_*(z)=1$ for all the possible $z(\mathrm{[Fe/H]})$ calculated using equation \ref{eq:zFe}.
\item \label{it:2} The MDF marginalised over $z$ for a single star is given by:
\beq
P_*(\mathrm{[Fe/H]})=\int_0^\infty Q(\mathrm{[Fe/H]}(z),z)P_*(z) dz
\eeq
where $Q(\mathrm{[Fe/H]}(z),z)$ is the observed MDF at each $z$ for all K dwarfs as described previously. 

\item \label{it:3} A new $P_*(z)$ is calculated through equation \ref{eq:pzFe}, using the $P_*(\mathrm{[Fe/H]})$ just computed. 

\item We restart steps \ref{it:2} to \ref{it:3} to calculate $P_*(\mathrm{[Fe/H]})$ with the new $P_*(z)$ until it converges. For all stars, we obtained convergence in less than 5 iterations.
\een

\begin{figure}
\centering
\begin{tabular}{c}
\includegraphics[width=0.5\textwidth]{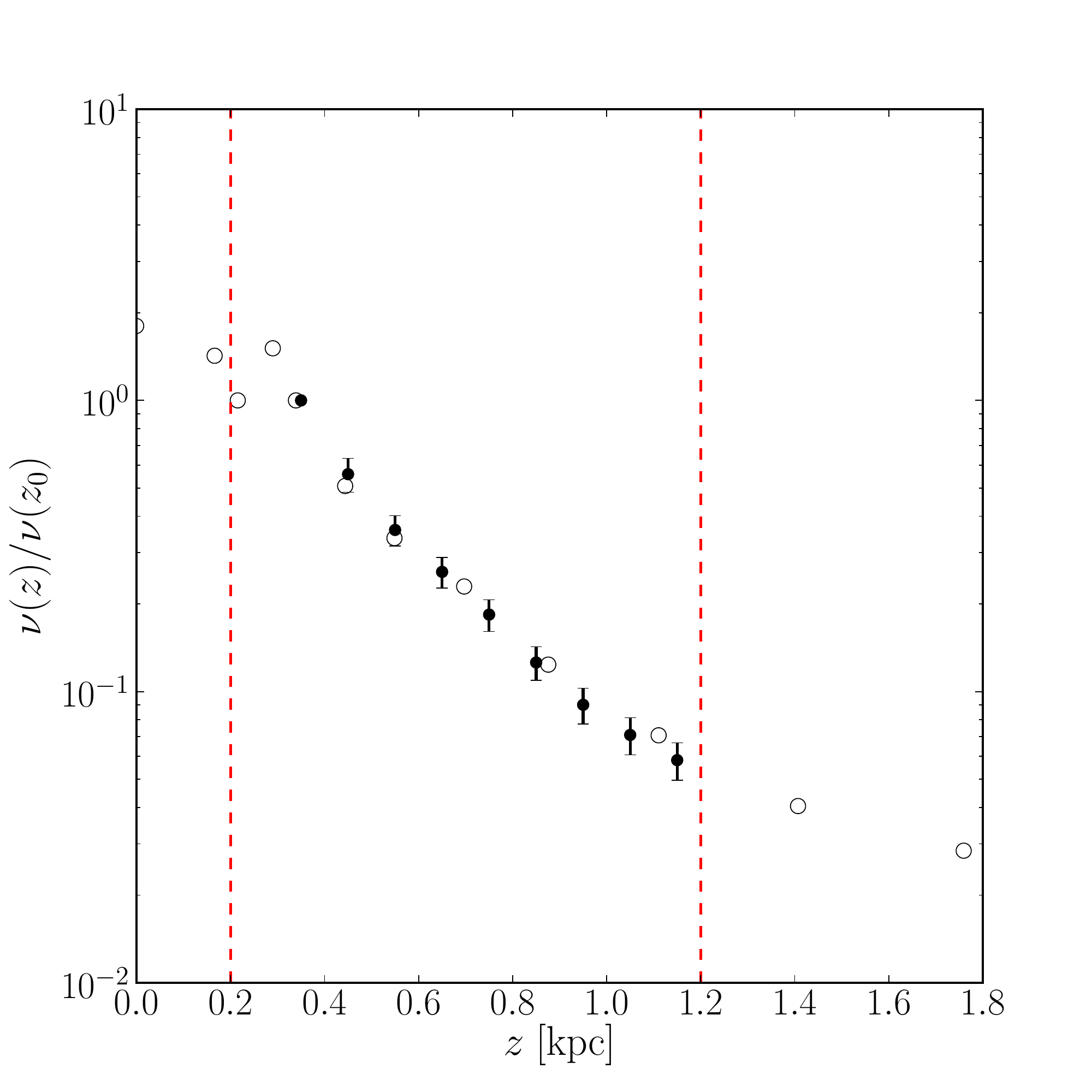}\\
\includegraphics[width=0.5\textwidth]{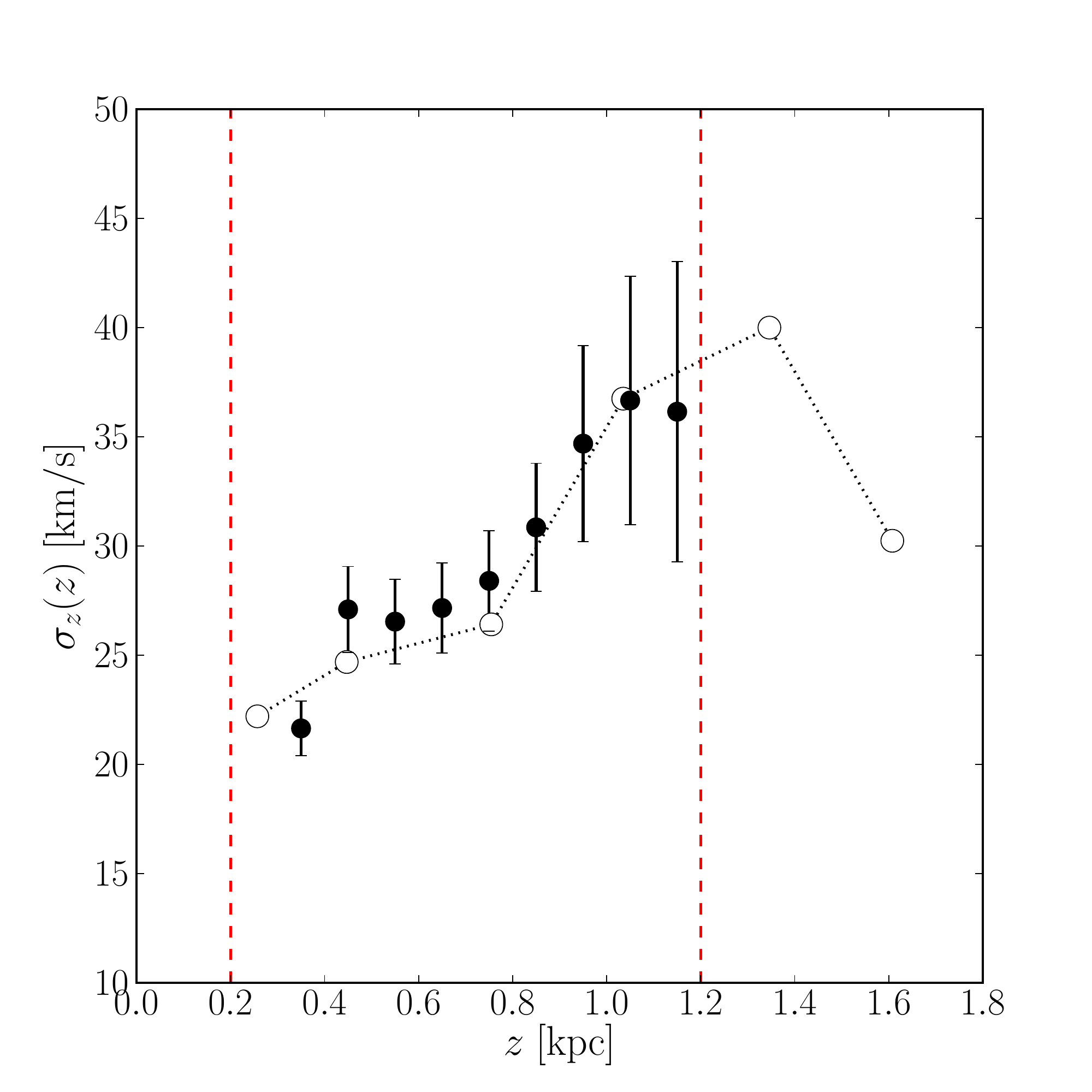}
\end{tabular}
\caption{{\it Upper panel:} K dwarf stellar density profile (filled circles with error bars) derived from a Monte Carlo sampling of $P_*(z)$ for each star. As a comparison, the density profile (assuming a metallicity gradient of $-0.3$\,dex/kpc) from KG89II is plotted as empty circles. {\it Lower panel}: The similarly derived vertical velocity dispersion as a function of $z$ (filled circles with error bars). The corresponding determination by KG89II is represented by the empty circles. In both panels, the two red dashed lines show the range of $z$ considered in our analysis, over which the photometric sample is volume complete and we avoid significant contamination from K giants stars.}
\label{fig:newdata}
\end{figure}

The density fall off of the photometric sample and the velocity dispersion of the spectroscopic one, obtained with our new distance estimates, are shown in figure \ref{fig:newdata}. For our analysis, we use the density and the velocity dispersion profiles only over the range $200<z<1200$\,pc (red dashed lines). This assures that our sample is volume complete and avoids significant contamination by K giant stars \cite{kg_1989b}. The corresponding quantities computed by KG89II, assuming a metallicity gradient of $-0.3$\,dex/kpc, are plotted as a comparison (empty circles). Our new density profile and velocity dispersion do not differ greatly from those of KG89II; they are compatible within the quoted errors. In Appendix \ref{app:met}, we explore the effect of a rather extreme variation in the assumed MDF -- ignoring \cite{kotoneva_2002} and SEGUE data -- finding that our results are not sensitive to plausible changes in our distance calibration.
The referee of this article pointed out that the study of \cite{kotoneva_2002} has been updated by \cite{casagrande2007}. The two studies are very much compatible, but the scatter in equation \ref{eq:poly} of 0.03\,mag becomes 0.27\,mag when the newer data are used. We tested the impact of this larger scatter in magnitude on the distance calibration, finding that the density and velocity dispersion profile remain unchanged in the range of $z$ of interest, with a negligible increase in the uncertainties (see Appendix \ref{app:met}).
 
\section{Method}\label{sec:Met}

\subsection{The MA method}\label{sec:MA} 
The MA method presented in Paper I uses the Poisson-Jeans system to predict the density fall off of a tracer population in a given gravitational potential. The comparison between this predicted density fall off and the observed one allows us to constrain the gravitational potential and, consequently, the underlying dark matter distribution. 

Here we summarise the basic equations; for a detailed description of the MA method see Section 2.1 of Paper I. 

The MA method is based on three main assumptions:
\ben
\item The system is in equilibrium (steady state assumption).\label{eq-hyp}
\item The dark matter density is constant over the range of $|z|$ considered. \label{dm-hyp}
\item The `tilt' term $\frac{1}{R}\frac{\partial}{\partial R}\left(R\nu \sigma^2_{Rz}\right)$ in the cylindrical Jeans equation:
\beq
\frac{1}{R}\frac{\partial}{\partial R}\left(R\nu \sigma_{Rz}^2\right)
  + \frac{\partial}{\partial z}\left(\nu_i \vztwoi \right) + \nu_i
  \frac{\partial \Phi}{\partial z} = 0
\label{eqn:jeans}
\eeq
is negligible compared to all other terms. Here $\nu$, $\vztwoi$ and $ \sigma^2_{Rz}$ are the number density and the velocity dispersion components of a tracer population moving in potential $\Phi$. \label{tilt-hyp}
\een
With these assumptions, the Jeans equation becomes a function only of $z$ and we can neglect the other two Jeans equations in $R$ and $\theta$:
\beq
 \vztwoi \frac{\partial \nu}{\partial z} + 
 \nu \left(\frac{\partial \Phi}{\partial z} +\frac{\partial  \vztwoi}{\partial z} \right)= 0. 
 \label{eqn:jeans2}
\eeq
Solving this equation for a single tracer population, we obtain its density $\nu(z)$ at each height $z$:
\beq
\frac{\nu(z)}{\nu(z_0)}=\frac{\vztwoi(z_0)}{\vztwoi(z)}\exp\left(-\int_{z_0}^{z}\frac{1}{\vztwoi(z)}\frac{d\Phi}{dz}dz\right)
\label{new_nu}
\eeq
Given the density at the midplane $\rho_{\mathrm{s},j}(0)$ and the vertical velocity dispersion $\vzstwoj(z)$ as a function of $z$ for each of the gas and stellar populations in the local disc, we can model the full disc density distribution as a superposition of such elements:
\beq
\rhos(z)=\sum_j \rho_{\mathrm{s},j}(0)\frac{\vzstwoj(0)}{\vzstwoj(z)}\exp\left(-\int_0^{z}\frac{1}{\vzstwoj}\frac{d\Phi}{dz}dz\right).
\label{eqn:new_rhodisc}
\eeq
In Paper I, we showed that accurate measurement of the vertical velocity dispersion of the tracers $\vztwoi(z)$ is crucial, however in the mass modelling we can assume that all the visible matter components are isothermal -- i.e. $\vzstwoj=\vzstwoj(0)$ -- and equation \ref{eqn:new_rhodisc} simplifies to:
\beq
\rhos(z)=\sum_j \rho_{\mathrm{s},j}(0)\exp\left(-\frac{\Phi(z)}{\vzstwoj(0)}\right).
\label{eqn:rhodisc}
\eeq
The Poisson equation then determines the potential $\Phi$ from the density. In cylindrical polar coordinates, this is given by: 
\begin{equation}
\frac{\partial^2\Phi}{\partial z^2}=4\pi G(\rhos(z)+\rhoeff)
\label{eqn:poisson}
\end{equation}
with:
\begin{equation}
\rhoeff=\rhodm(R)-(4\pi G R)^{-1}\frac{\partial}{\partial R}V^2_c(R)
\label{dmeff}
\end{equation}
where $\rhodm(R)$ is the halo mass density (following assumption \ref{dm-hyp}, this is assumed to be independent of $z$ in the volume considered); and $V_c(R)=(R\partial\Phi/\partial R)^{1/2}$ is the (total) circular velocity at a distance R (in the plane) from the centre of the Galaxy. For a flat rotation curve, the second term vanishes and $\rhoeff(R)=\rhodm(R)$. The rotation curve correction can be calculated from the Oort constants $A$ and $B$ \citep{BinneyMerri}:
\beq
(4\pi G R)^{-1}\frac{\partial V^2_c}{\partial R}=\frac{B^2-A^2}{2\pi G}
\eeq

We solve equations \ref{eqn:rhodisc} and \ref{eqn:poisson} numerically for a given tracer population, adopting the following procedure: 
\ben 
\item We make initial trial guesses for $\rho_{\mathrm{s},j}(0)$, $\rhodm$, and the vertical velocity dispersion for the visible matter component in the plane $\vzstwoj(0)$.
\item We solve equation \ref{eqn:rhodisc} to obtain $\Phi(z)$ and its first derivative $\frac{\partial\Phi}{\partial z}$, with $\Phi(0)=\left.\frac{\partial\Phi}{\partial z}\right|_{0}=0$. 
\item We insert this result into equation \ref{new_nu} for the vertical density fall off of the tracers $\nu_\mathrm{p}(z)$ and we compare this with the observed distribution $\nu(z)$ to obtain a goodness of fit. 
\een
This procedure requires many input parameters, such as the normalisations and dispersions of each of the disc components, and the vertical dispersion profile of the tracers (typically poorly constrained). To explore the parameter space and marginalise over the uncertainties, we use a Monte Carlo Markov Chain (MCMC) method. 

In Paper I we showed that our method is able to recover the correct value of the local dark matter density, even in presence of large visible matter density fluctuations due to the spiral arms. Notice that the MA requires no prior from the Milky Way rotation curve, as has been commonly used in previous works; this means that we can compare our determination to that derived from the rotation curve to constrain the Milky Way halo shape.

\subsubsection{Application of the MA method to real data}\label{sec:apply}
When we apply the MA method to real data, we must deal with distance and velocity uncertainties, and account for survey geometry and/or the sample completeness. In particular, the K dwarf data from KG89II are not assigned distances, but instead, as described in Section \ref{sec:newdist}, a $z$ probability distribution function $P_*(z)$ for each star of the sample. In addition, the vertical velocities for each star are measured with an uncertainty of about $0.5-1$\,km/s. To marginalise over these uncertainties, we proceed in the following way:
\ben
\item\label{it:beg} For each model $n$ of the MCMC, we select a different vertical distance $z^*_{n}$ for each star in the sample, according to its $z$ distribution function $P_*(z)$. We make sure that for each star included in both the spectroscopic and the photometric sample we pick a unique $z^*_{n}$ value. For each star of the spectroscopic sample we also draw a vertical velocity value $v^{*}_{z,n}$ from a Gaussian distribution, according to its velocity error bar.
\item \label{it:bin}  We bin the data in $z$ to construct the observed density fall off $\nu_n(z)$ and velocity dispersion $\vztwon(z)$ for the tracers, selecting stars with $z$ between $0.2$ and $1.2$\,kpc. The velocity dispersion is calculated using the velocity scale algorithm described in \cite{biweight}.
\item We use the trial guesses for $\rho_{\mathrm{s},j,n}(0)$, $\vzstwojn(0)$ and $\rho_{\mathrm{dm},n}$ to solve equation \ref{eqn:rhodisc} and \ref{eqn:poisson} to obtain $\Phi_n(z)$ and its first derivative $\frac{\partial\Phi_n}{\partial z}$.
\item We insert this result and the vertical velocity dispersion of the tracers $\vztwon(z)$ into equation \ref{new_nu} to predict the tracers' density fall off. When applying the MA method to the real data, we add the visible matter surface density $\Sigma_\mathrm{s}$ as a further constraint. For each model of the MCMC we compute $\Sigma_\mathrm{s}$ as
\beq
\Sigma_{\mathrm{s},n}=2\int_0^{\infty}\rhos(z)dz=\int_0^\infty\sum_j \rho_{\mathrm{s},j,n}(0)\exp\left(-\frac{\Phi_n(z)}{\sigma^2_{z,j,n}}\right)\label{eqn:Sigma}
\eeq

\item We compare the predicted density fall off $\nu_{n,\mathrm{p}}(z)$ with $\nu_n(z)$, the predicted visible matter surface density $\Sigma_{\mathrm{s},n}^\mathrm{p}$ with the observed one ($\Sigma_{\mathrm{s}}$) and we calculate the corresponding $\chi^2_n$, accepting or rejecting the model $n$.
\item When a model is accepted, we restart from \ref{it:beg} with the following  model ($n+1$) and so on, exploring the whole parameter space.
\een


\subsection{The KG method} \label{sec:KG}

KG89II used the K dwarf data to calculate the total surface density at the Sun position $\Sigma(R_\odot)$. Their approach (the KG method) is similar to the method adopted later by \cite{holmberg_local_2000,holmberg_local_2004} (the HF method), that we analysed in detail in Paper I.

Instead of measuring the vertical velocity dispersion of the tracers as a function of $z$ to predict the density fall off, the HF method uses the tracers' velocity distribution function in the mid-plane of the Galactic disc $f_z(v_{z,0})$; assuming that the distribution function is separable, $f=f_{R,\theta}(v_R,v_\theta,R)\times f_z(v_z,z)$. \cite{holmberg_local_2000,holmberg_local_2004} integrate this distribution function over $z$-velocities to predict the density fall off of the tracers (see Section 2.2 of Paper I for more details):
\beq
\nu(z)=\int_{-\infty}^\infty f_z(v_z,z)dv_z=2\int_{\Phi(z)}^\infty\frac{f_z(E_z)dE_z}{\sqrt{2[E_z-\Phi(z)]}}\label{eqn:nuHF}
\eeq
where $f_z(z,v_z)=f_z(E_z)$ and $E_z=\frac{1}{2}v_z^2+\Phi(z)$ is the vertical energy. This equation can be written as:
\beq
\nu(z)=2\int_{\sqrt{2\Phi(z)}}^\infty\frac{f_z(v_{z,0})v_{z,0}dv_{z,0}}{\sqrt{v_{z,0}^2-2\Phi(z)}}\label{eqn:nuHF2}
\eeq
where $v_{z,0}$ is the vertical velocity at the Galactic mid-plane ($z=0$). 

The MA demands only that the tilt term in the Jeans equation (\ref{eqn:jeans}) is {\it small} with respect to the other terms, the HF method requires the stronger assumption that the $z$-motion (and so the distribution function) is completely separable from the motion in the radial and azimuthal directions; this latter implies that the tilt term is {\it exactly zero}.

The HF approach has the advantage of exploiting the whole available information about the shape of the velocity distribution function of the tracers. However, we demonstrated in Paper I that when the separability of the distribution function is not fulfilled, the HF method leads to biased results. Using a high resolution simulation of a Milky Way like galaxy, we showed that the onset of spiral arms and a bar can cause significant radial mixing that breaks the separability of the motion in the $z$ and $R$ directions, violating this key assumption of the HF method. This effect becomes increasingly important with height $z$. It is important to notice that the HF method can not be corrected for this bias since the separability of the potential (and of the distribution function) lies at the heart of the method. By contrast, in the MA method, the separability of the potential enters only in the neglected tilt term of the Jeans equation (that is assumed to be small as compared to the other terms). If the tilt term is for some reason large -- i.e. the radial derivative of the density weighted tilt of the velocity ellipsoid is large -- a correction can straightforwardly be applied to our MA method. However, we expect the tilt term to be small \citep[see Paper I and][]{binney_galactic_2008}, and so we do not consider such a correction in this paper.

KG89II's approach relies on the same key assumption about the distribution function as the HF method. In most studies, the density $\nu(z)$ of the tracers is known to better precision than the velocity distribution $f_z(v_z,z)$. For this reason, KG89II work in the opposite direction with respect to \cite{holmberg_local_2000,holmberg_local_2004} and predict $f_z(v_z,z)$ from the observed $\nu(z)$. Applying an inverse Abel transform to equation \ref{eqn:nuHF}, they obtain:
\beq
f_z(E_z)=\frac{1}{\pi}\int_{E_z}^\infty\frac{-d\nu/d\Phi}{\sqrt{2(\Phi-E_z)}}d\Phi. \label{eq:fEz}
\eeq
so there is a unique relation between $\nu(\Phi)$ and $f_z(E_z)$. Notice that $f_z(E_z)$ depends on $\nu(\Phi(z))$ only at large $z$, where the potential exceeds $E_z$, i.e. beyond $z=\Phi^{-1}(E_z)$. Thus an additional key advantage of the KG method is that one can model the potential at large distances from the Galactic plane, ignoring the detailed distribution of matter at small $z$. KG89II parameterised the gravitational potential $\Phi(z)$ above the bulk of the disc matter (where it is sensitive only to the total surface density of gravitating matter) as:
\beq
\Phi(z)=K(\sqrt{z^2+D^2}-D)+Fz^2
\eeq
where $D$ is the disc scale height, $K$ is proportional to the total disc surface density $\Sigma(R_\odot)$, and $F\propto \rho_\mathrm{dm}^\mathrm{eff}$ (the effective halo density). KG89II used a range of Galactic mass models (calculated using different values of the disc mass $M$, the radial disc scale-length $R_\mathrm{d}$, the circular velocity $V_c(R_\odot)$ and Sun position $R_\odot$) to {\it ensure} consistency with the Galactic rotation curve (assuming a spherical Milky Way halo) and therefore to obtain a relation between $F$ and $K$. Note that already this is different from our MA approach where we use no information about the rotation curve to constrain our mass models.

Given the observed space density of a tracer population $\nu(z)$ and a set of gravitational potential models $\Phi(z)$, one can solve equation \ref{eq:fEz}. To reduce the noise in the differential of $\nu(z)$, KG89II fitted it with a double exponential. KG89II then used the derived $f_z(E_z)$ for each potential model $\Phi(z)$ to compute the likelihood of the spectroscopic sample:

\beq
\mathcal{L}=\sideset{}{_*}\prod \frac{f_z(E_{z,*})}{\int_0^\infty f_z(E_z)dE_z}\label{eqn:LIK}
\eeq
where the product is over all stars in the spectroscopic sample, and select the potential parameters that maximise this likelihood function $\mathcal{L}$.

The KG method, like the HF method, uses the full shape of the observed velocity distribution function, maximising the use of the available information. It is also convenient because it does not require a detailed model of the gravitational potential or an accurately measured tracer density fall off close to the Galactic plane. However, its drawback is that, like the HF method, it relies on a key assumption that the vertical distribution function is only a function of $E_z$. In the following section we test, using the high resolution N-body simulation described in Paper I, how this assumption affects the result derived using the KG method.

\subsection{Testing the methods using an N-body simulation}\label{sec:Nbody}


\begin{figure*}
\centering
\begin{tabular}{cc}
\includegraphics[width=0.38\textwidth]{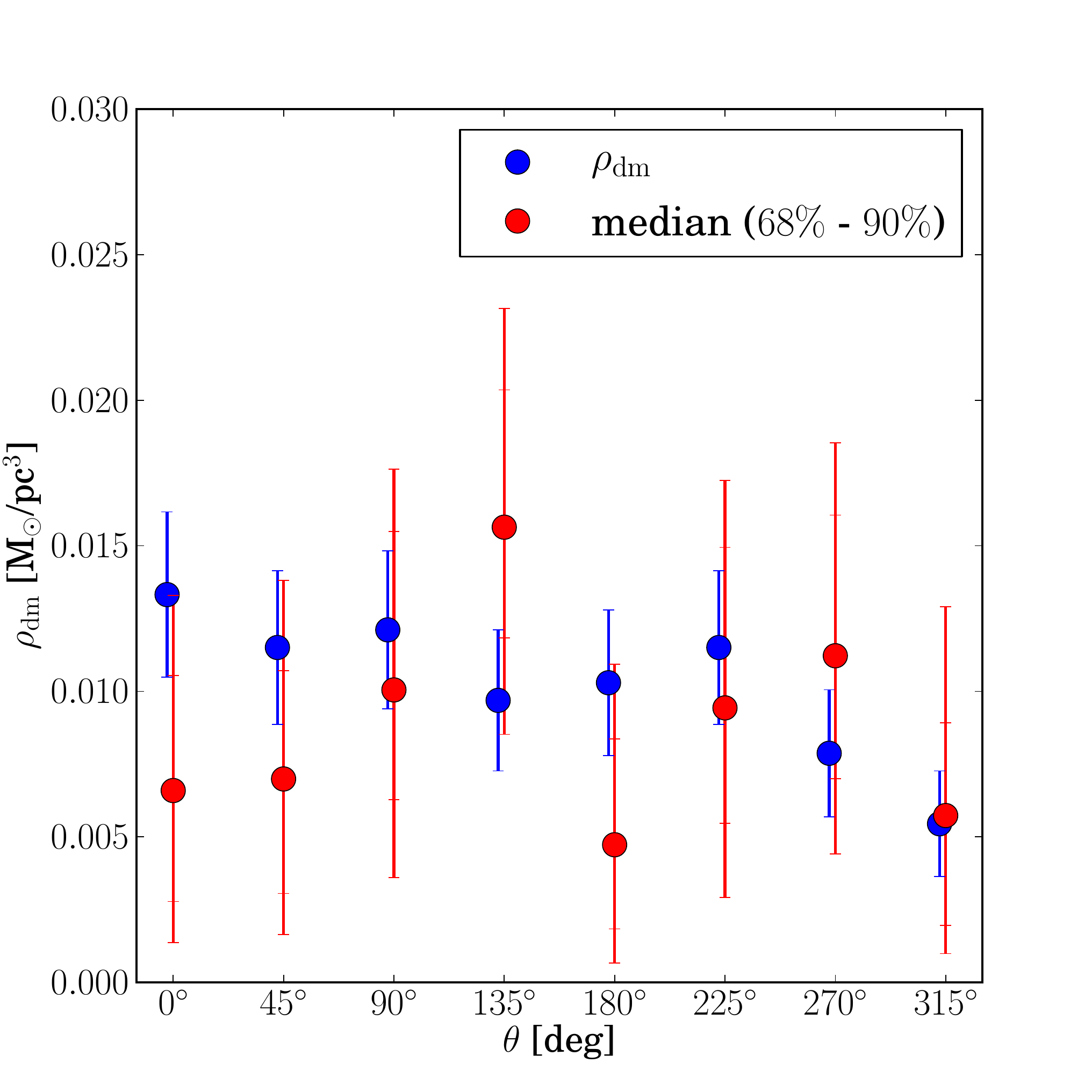} &
\includegraphics[width=0.38\textwidth]{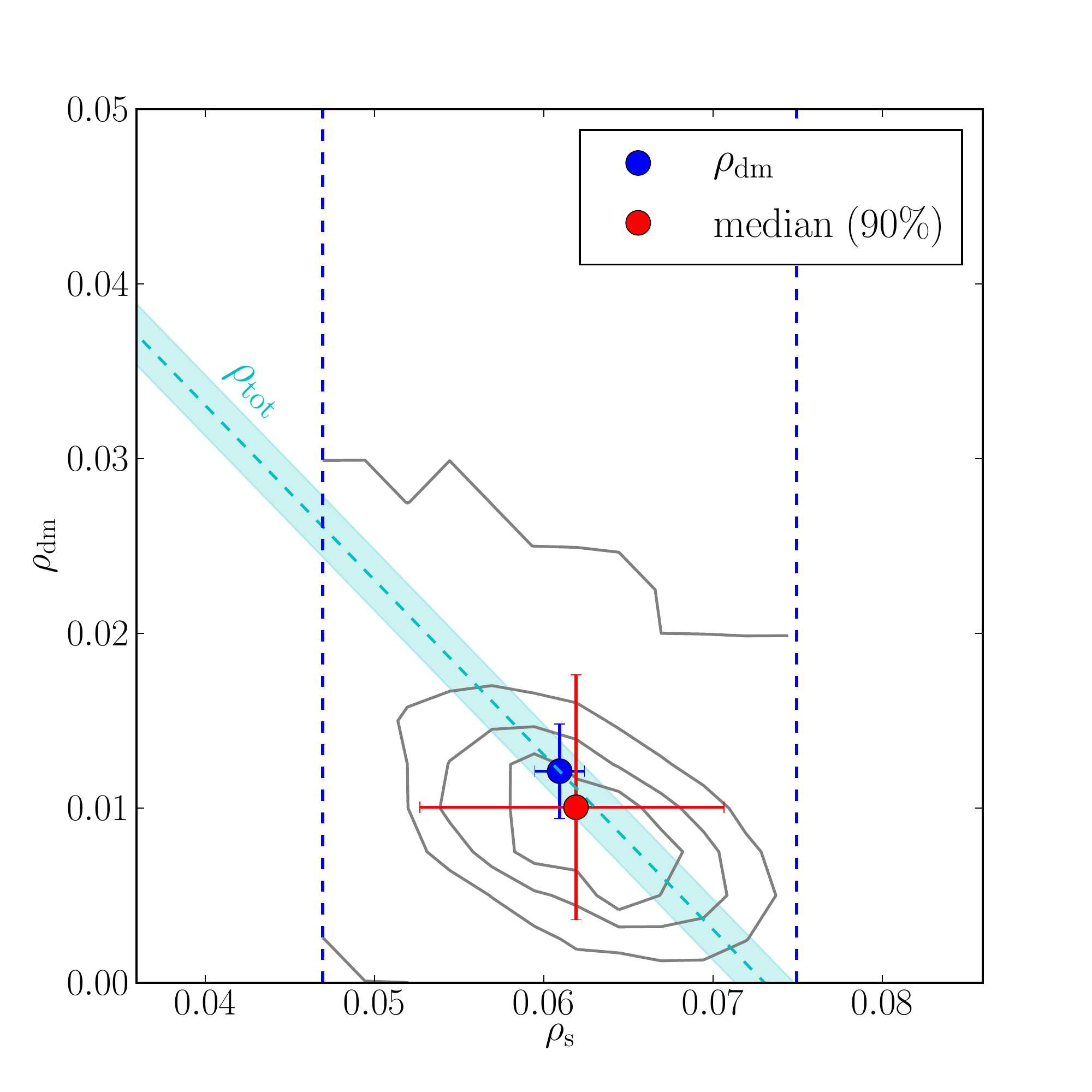} \\
\includegraphics[width=0.38\textwidth]{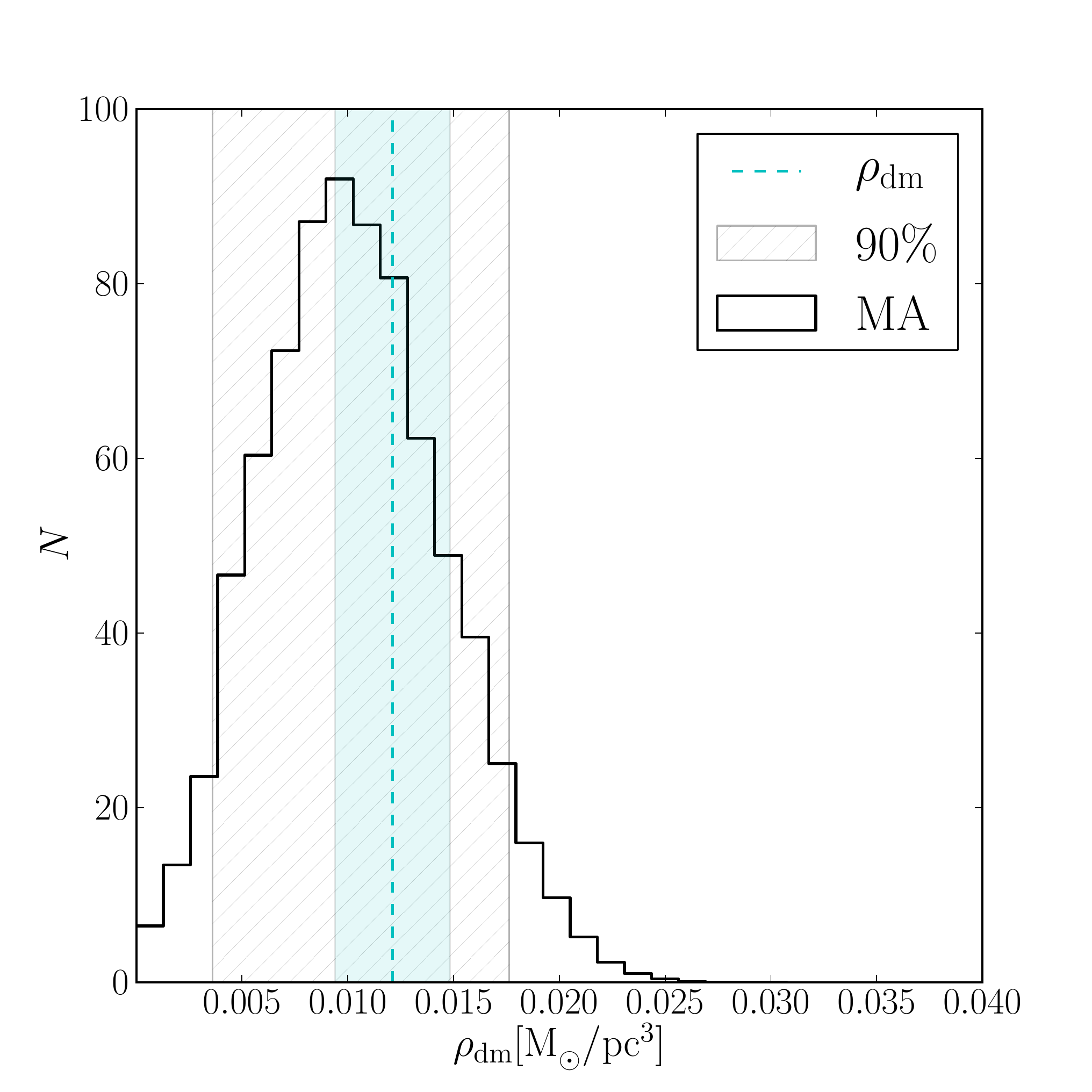} &
\includegraphics[width=0.38\textwidth]{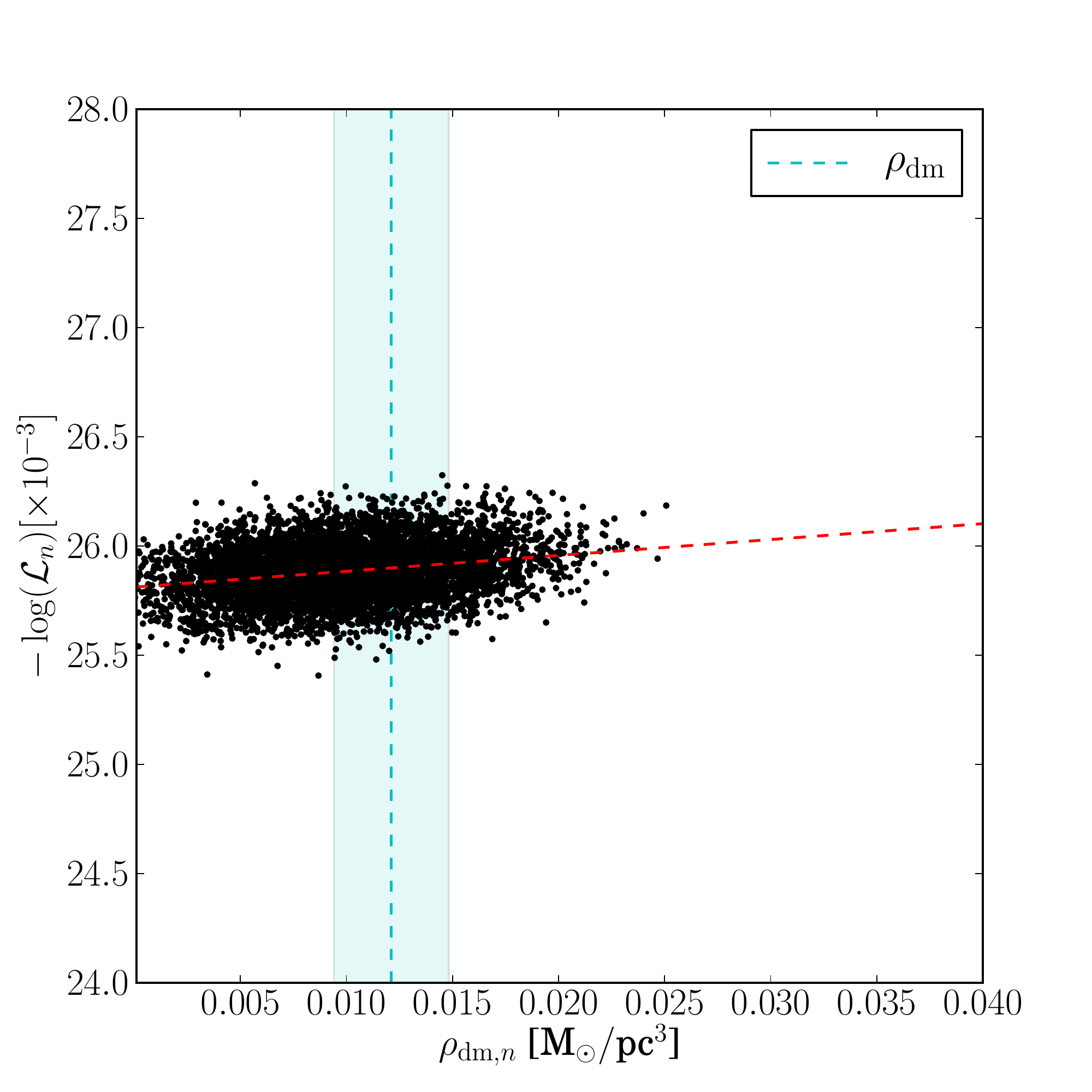}
\end{tabular}
\caption{Results for recovering $\rhodm$ and $\rhos$ from the simulation. {\it Top left panel}: The recovered dark matter density (red filled circles) for all 8 volumes analysed around the disc (90\% and 68\% confidence intervals are shown). The actual value of the dark matter density is marked as a blue filled circle. {\it Top right panel}: The density of models explored by the MCMC projected onto $\rhodm$-$\rhos$ space for the 90$^\circ$ volume. The blue dot shows the true value for $\rhodm$ and $\rhos$; the diagonal cyan blue line shows the total matter density; and the red dot shows the median recovered $\rhodm$ and $\rhos$, with 90\% errors marked. {\it Bottom left panel}: Histogram of the recovered $\rhodm$ from the MCMC mode ensemble for this same volume. The striped grey area is the $90\%$ confidence interval. The cyan dashed line and shaded area give the actual value of $\rhodm$ with error bars. {\it Bottom right panel:} Testing the effect of assuming a separable distribution function. Each dot in this plot shows the likelihood $L_n$ of each MCMC model calculated assuming a separable distribution function (see equation \ref{eq:fEz}; the plot shows $-\log(\mathcal{L}_n)$, so the more likely models have a {\it lower} ordinate value). The red dashed line shows a fit to the points. Notice that assuming that the distribution function is separable produces a bias towards low $\rhodm$ (the true distribution function for the simulation is not separable; see figure \ref{fig:fEz}). This bias effect was observed in all the volumes considered.}
\label{fig:KGvsMA}
\end{figure*}

Before applying the MA method to the real K dwarf data, we use the most dynamically evolved stage of the simulation described in Paper I as a mock data set, to test the effect of the velocity errors and of the asymmetric distance distribution function $P_*(z)$ on the MA method's result. Then we use the same mock data to probe how the non-separability of the tracers' distribution function affects the KG method. 

The mock data consist of a high resolution ($30$ million disc star particles and $15$ million halo dark matter particles) N-body simulation of an isolated Milky Way like galaxy. The initial conditions were built to contain some thousand stars in the volume size required for our analysis. With the dynamical evolution of the simulation, the disc developed a bar and spiral arms. For more details about the simulation features and how it compares to the real Milky Way, see Section 3.1 of Paper I. 

We consider several volumes in the disc of the simulated galaxy at a distance of $R_\odot=8.5$\,kpc from the centre. The MA method solves the Jeans equation for a one-dimentional slab (equation \ref{eqn:jeans2}), so the radial size of the volumes, $\Delta R=0.25$\,kpc, is chosen to fulfil this approximation, but still contain an enough large number of stars (see Section 3.3.1.1 of Paper I for more details).

As described in Section \ref{sec:apply}, we assign a different velocity value $v^*_{z,n}$ and a different $z_n^*$ to each star at every iteration $n$ of the MCMC. In the application of the MA method to the simulation, the velocity values are drawn from a Gaussian distribution centred on the true velocity value and with a width of $1$\,km/s, while the $z$ values are selected from a lognormal distribution around the true value. Because of the numerical resolution of the simulation, we cannot fit the density profile up to $1.2$\,kpc, since at such height we quickly run out of star particles and the velocity dispersion is poorly measured, so we use stars with $0.2\leq z \leq0.75$\,kpc. For the simulation data, we model the visible mass distribution as a single component, characterised by its mass density $\rho_{\mathrm{s},j}=\rhos(0)$ and its velocity dispersion on the midplane $\vzstwoj=\vztwo(0)$. We let the dark matter density freely vary between $0$ and $0.2$\,\msun\,pc$^{-3}$, and the other parameters --  $\rhos(0)$ and $\vztwo(0)$ -- vary inside their error bars. We adopted Poisson errors for the velocity dispersion $\vztwo(0)$ and the current uncertainties on the total visible matter density in the plane (i.e. $0.014$\,\msun\,pc$^{-3}$, see Section \ref{sec:rhodmdata}) for $\rhos(0)$. We include the rotation curve correction term, which can be easily computed for the simulation, in the calculation. We test convergence of our MCMC chains by starting with $\rhodm$ seeded at two different values (namely $\rhodm = 0$ and $\rhodm = 0.2$\msun\,pc$^{-3}$) and running until the two chains are statistically indistinguishable.

In figure \ref{fig:KGvsMA}, the results for the MA method are shown. The upper left panel shows the results for eight different volumes around the simulated disc. Notice that in all cases, the mean correct answer (blue filled circles) is recovered within the 90\% confidence interval, while for four out of eight of the patches (with a fifth at 45$^\circ$ extremely close) it is recovered with the 68\% confidence interval. This is consistent with our confidence intervals having the meaning of a purely statistical error, despite each patch being systematically different (each patch samples a different region of the disc with different local dynamics). The remaining three panels focus on the results for the volume at 90$^\circ$. The top right panel shows the density of models explored by the MCMC projected onto $\rhodm$-$\rhos$ space; the bottom left panel shows a histogram of the dark matter density for all the models explored by the MCMC; and the bottom right panel explores the effect of assuming a separable distribution function (of which more, next).

In Paper I, we noticed that, at this evolved stage of the simulation, the distribution function $f(v_R,v_\theta,v_z,R,z)$ of the tracers is not separable. This leads to the HF method producing biased results -- either underestimating or overestimating the local dark matter density (see Section 3 of Paper I for more details). 

The separability of the distribution function lies at the heart of the KG method too. To reproduce a KG-like method, we consider all $\nu_n(z)$ and model potentials $\Phi_n(z)$ explored by our MA method MCMC chain. For each model in the chain, we then calculate a distribution function $f_{z,n}(E_z(v_z,\Phi))$ through equation \ref{eq:fEz} and use it to compute the likelihood $\mathcal{L}_n$ of the velocity data via equation \ref{eqn:LIK}.

\begin{figure}
\centering
\begin{tabular}{c}
\includegraphics[width=0.5\textwidth]{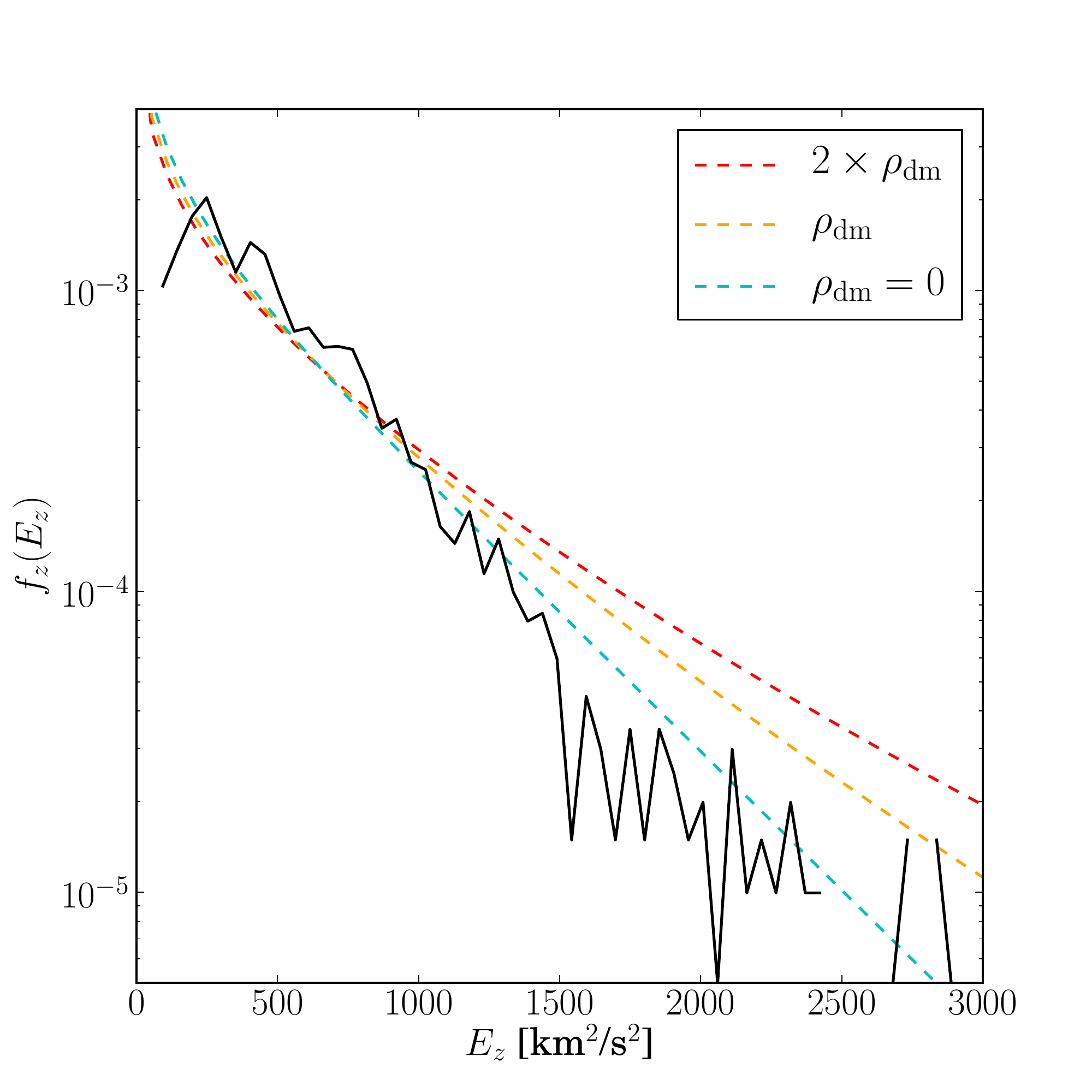}\\
\end{tabular}
\caption{The distribution function -- $f$ -- of the tracers in the simulation. The black line shows distribution function measured directly from the simulation averaged over a volume $8.375 < R < 8.625$\,kpc; $0.2 < z < 0.75$\,kpc. The orange dashed line shows $f_z(E_z)$ predicted from the density fall off $\nu(z)$ via equation \ref{eq:fEz}, assuming the correct value of $\rhos$ and $\rhodm$ in the computation of the gravitational potential. The blue and red dashed lines show $f_z(E_z)$ predicted from $\nu(z)$, assuming $\rhodm=0$ and twice $\rhodm$, respectively. All distribution functions are normalised by the integral of $f_z(E_z)$ between the minimum and the maximum $E_z$ of the star particles in the volume considered.}
\label{fig:fEz}
\end{figure}

In practice, the integral in the denominator of equation \ref{eqn:LIK} is calculated numerically from $E^\mathrm{min}_z=1$ to $E^\mathrm{max}_z=7000$\,km$^2$\,s$^{-2}$, which is chosen to avoid the divergence at $E_z=0$ and to ensure we cover all energies of interest (the contribution of the high energy tail of $f_{z,n}(E_z)$ is negligible with respect to the low energy part; see figure \ref{fig:fEz}). 

In the bottom right panel of figure \ref{fig:KGvsMA}, the likelihoods $\mathcal{L}_n$ of the velocity data are plotted against the corresponding values of the local dark matter density for the different MCMC models $\rho_{\mathrm{dm},n}$. This panel shows that there is an anti-correlation between the computed likelihood $\mathcal{L}_n$ and the corresponding value of the local dark matter density $\rho_{\mathrm{dm},n}$: the likelihood of the velocity data is larger (i.e. $-\log (\mathcal{L}_n)$ is lower) for gravitational potential models with low $\rho_{\mathrm{dm},n}$; this means that we expect the KG method to artificially favour low dark matter density values. For all the explored volumes around the disc, we always obtain this same anti-correlation, so the bias on $\rhodm$ will have always the same sign. 

To understand this effect, in figure \ref{fig:fEz} we show the distribution function $f_z(E_z)$ calculated from the same density profile $\nu(z)$, but using three potentials with different values of $\rhodm$, namely $\rhodm=0$ (blue dashed line), the true value of $\rhodm$ (orange dashed line) and twice the true $\rhodm$ (red dashed line); the black line represents the actual distribution function of the stars in the volume. The potential corresponding to the true value of the dark matter density does not predict the distribution function correctly, while with $\rhodm=0$ we obtain a better agreement with the measured $f_z(E_z)$. It is clear from this plot that the likelihood of models with low dark matter density, calculated through equation \ref{eqn:LIK}, will be higher than the true model. Therefore the KG method will be biased towards low $\rhodm$.

\section{Results}\label{sec:Res}

\subsection{Measuring the local matter and dark matter density} \label{sec:rhodmdata}
In the previous section, we showed that the MA method is able to recover the correct dark matter density within our 90\% confidence interval, even in presence of asymmetric distance errors, velocity uncertainties and a non-separable distribution function. We now apply the MA method to the real K dwarf data, proceeding as described in Section \ref{sec:apply}. The mass distribution in the Galactic disc is modelled as a superposition of 15 isothermal components, listed in Table \ref{mmodel}. As parameters to fit in the MCMC, we use the local dark matter density $\rhodm$, the total visible density in the midplane $\rhos(0)$, and the relative fractions of the visible components $\rho_{\mathrm{s},j}(0)/\rhos(0)$ and their velocity dispersions in the midplane $\sigma_{z,j}$. We allow the densities and the velocity dispersions of the different components to vary within their measured uncertainties (the errors for each component are given in Table \ref{mmodel}). We let the total visible density in the plane $\rhos(0)$ vary within its observed range: $\rhos(0)=0.0914\pm0.0140$\msun\,pc$^{-3}$ (extrapolated from Table \ref{mmodel}); and we let the dark matter density vary between 0 and 0.2\msun\,pc$^{-3}$. For each model explored by the MCMC, we calculated the visible surface density $\Sigma^\mathrm{p}_{\mathrm{s,}n}$ through equation \ref{eqn:Sigma} and compare it with the total surface density we obtain from Table \ref{mmodel}, i.e. $\Sigma^\mathrm{obs}_\mathrm{s}\pm \Delta\Sigma^\mathrm{obs}_\mathrm{s}=49.4\pm 4.6$\,\msun\,pc$^{-2}$, to calculate the total $\chi^2$: 

\begin{equation} 
\chi^2 = \chi^2_\mathrm{surf} + \chi^2_\nu 
\end{equation} 
where:
\beq
\chi^2_\mathrm{surf}=\frac{ (\Sigma^\mathrm{obs}_\mathrm{s}-\Sigma^\mathrm{p}_{\mathrm{s},n})^2}{(\Delta\Sigma^\mathrm{obs}_\mathrm{s})^2}
\eeq
and
\beq
\chi^2_\nu=\sum^9_{i=1} \frac{(\nu_{n,i}-\nu_{n,\mathrm{p},i})^2}{(\Delta\nu_{n,i})^2},
\eeq
where the sum is extended to all the bins and $\Delta\nu_{n,i}$ are the uncertainties on the density fall off.

The rotation curve correction term can be calculated from the Oort constants $A$ and $B$. To determine the Oort constants, we must use stellar tracers that are well-mixed. The most recent estimates of $A$ and $B$ from F giants \citep{branham_oortF_2010} and K-M giants \citep{mignart_oorthipp_2000} from {\it Hipparcos} give $A=14.85 \pm 7.47$\,km\,s$^{-1}$\,kpc$^{-1}$ and $B=-10.85\pm 6.83$\,km\,s$^{-1}$\,kpc$^{-1}$ and $A = 14.5\pm 1.0$\,km\,s$^{-1}$\,kpc$^{-1}$ and $B =-11.5 \pm 1.0$\,km\,s$^{-1}$\,kpc$^{-1}$, respectively.  Averaging these two values we obtain a correction term of $-0.0033\pm0.0050$\msun\,pc$^{-3}$. We test for convergence of the MCMC by starting several chains at different initial values of all the parameters and running until they are statistically equivalent (after removing an initial burn-in phase of 100 accepted models for each chain).

\begin{table}
\center
\caption{The disc mass model taken from Flynn et al. 2006. For each component in the table, we give the local mass density in the midplane $\rho(0)$ in \msun pc$^{-3}$; the total column density $\Sigma$ in \msun pc$^{-2}$; and the vertical velocity dispersion $\sigma_{z,j}(0)$ in km\,s$^{-1}$. Uncertainties on the densities are assumed to be $50$\% for all the gas components (indicated with $^*$) and $20$\% for all of the stellar components. The largest uncertainties come from the gas that remains poorly constrained (compare, for example, compilations in Flynn et al. 2006, \protect\cite{binney_merrifield} and \protect\cite{ferriere2001}). For the thick disc, the column density is well known, while the velocity dispersion and the volume density are poorly known such that they should have larger error bars. However, these two quantities are essentially nuisance parameters for our analysis here. Since they anti-correlate and -- as pointed out by \protect\cite{kg_1989a} -- the local gravitational potential is mainly constrained by the column density, we simply assume small errors for both here such that the integrated column agrees with the observed value.}
\label{mmodel}
\begin{tabular}{|c|c|c|c|}
\hline
Component &$\nu_{i,0}(0)$& $\Sigma_i$ &  $\vztwoi(0)^{1/2}$ \\
& [\msun\,pc$^{-3}$] & [\msun pc$^{-2}$] & [km\,s$^{-1}$] \\
\hline
H$_2^*$ & 0.021 & 3.0 & $4.0\pm 1.0$ \\
HI(1)$^*$ & 0.016 & 4.1& $7.0\pm 1.0$ \\
HI(2)$^*$  & 0.012  & 4.1& $9.0\pm1.0$ \\
Warm gas$^*$ & 0.0009 & 2.0 & $40.0\pm1.0$\\
Giants & 0.0006  & 0.4 & $20.0\pm 2.0$ \\
$M_V<2.5$ & 0.0031 &0.9 &$7.5\pm 2.0$ \\
$2.5 < M_V < 3.0$ &  0.0015  & 0.6 & $10.5\pm 2.0$ \\
$3.0 < M_V < 4.0$ &  0.0020  & 1.1 & $14.0\pm 2.0$  \\
$4.0 < M_V < 5.0$ &  0.0022 & 1.7 & $18.0\pm 2.0$  \\
$5.0 < M_V < 8.0$ &  0.007 & 5.7 & $18.5\pm 2.0$\\
$M_V > 8.0$ & 0.0135  & 10.9&$18.5\pm 2.0$ \\
White dwarfs &	0.006  & 5.4&$20.0\pm 5.0$ \\
Brown dwarfs & 0.002 & 1.8&$20.0\pm 5.0$ \\
Thick disc& 0.0035  & 7.0&$37.0\pm 5.0$  \\
Stellar halo & 0.0001 & 0.6&$100.0\pm 10.0$  \\
\hline
\end{tabular}
\end{table}

\begin{figure}
\centering
\begin{tabular}{c}
\includegraphics[width=0.5\textwidth]{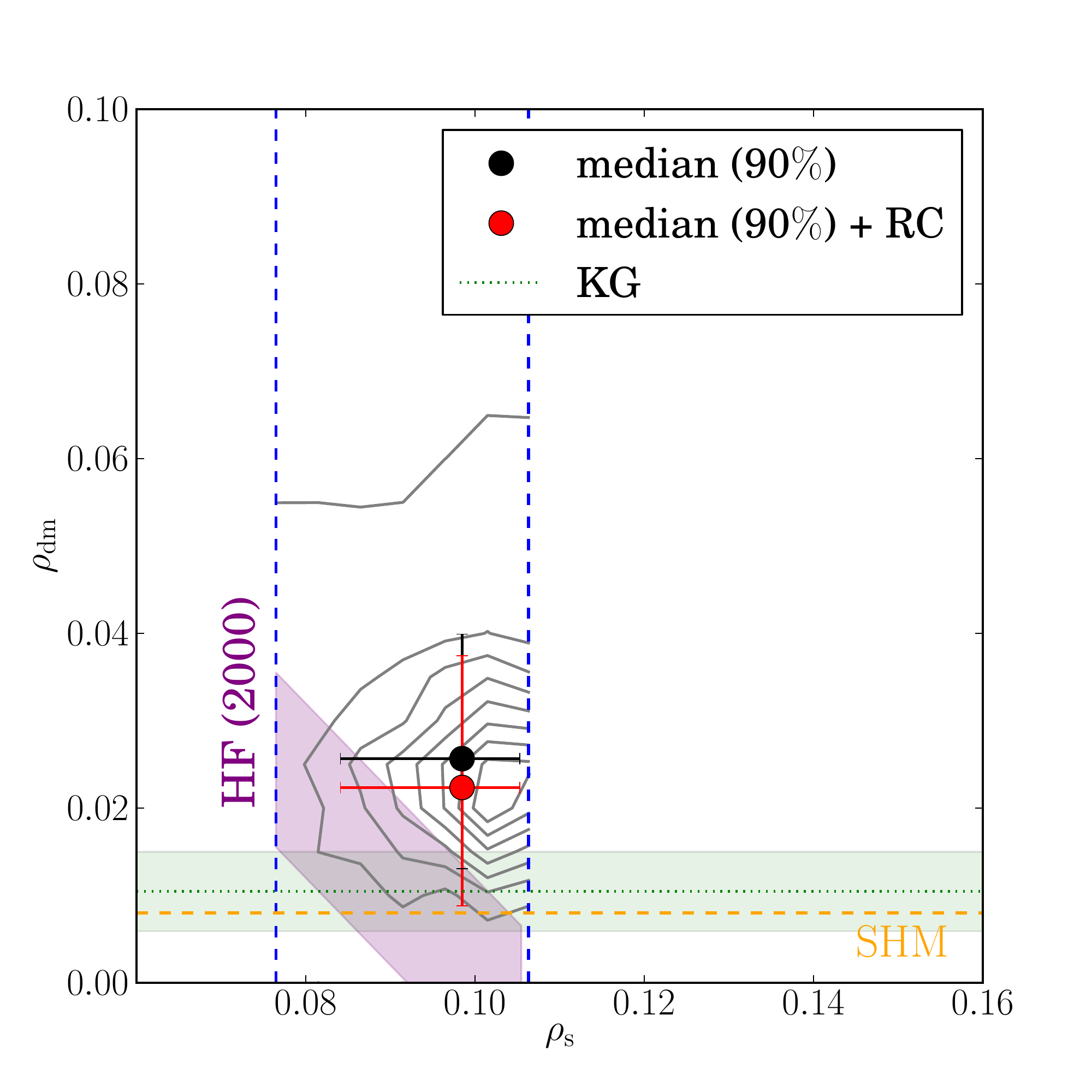}\\
\includegraphics[width=0.5\textwidth]{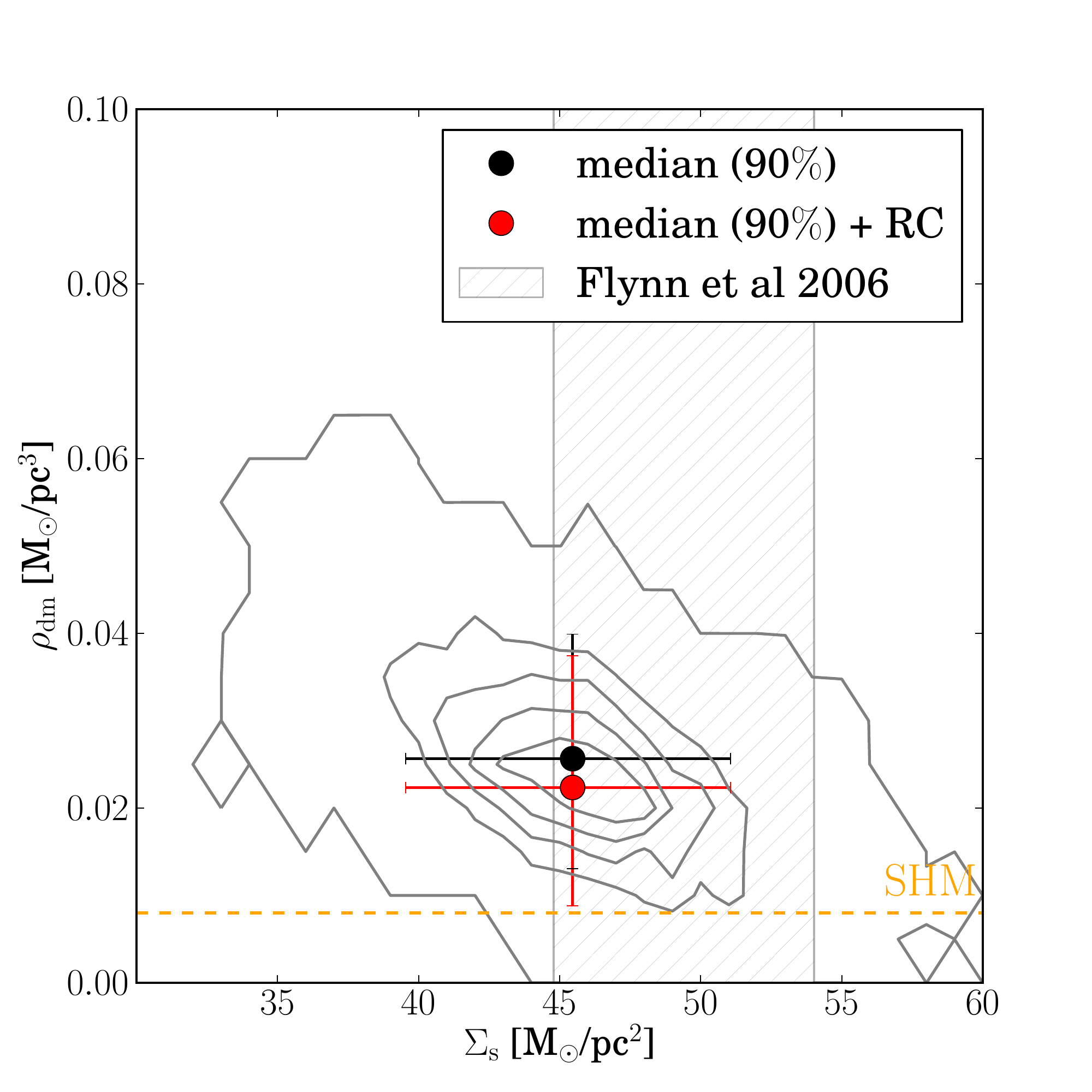}
\end{tabular}
\caption{{\it Upper Panel:} The recovered visible and dark matter densities. The grey contours are the density of models explored by the MCMC. The black dot shows the median recovered value of $\rhos$ and $\rhodm$ with 90\% errors marked; the red dot shows the same but including a correction for the local non-flatness of the Milky Way rotation curve ($-0.0033\pm0.0050$\,\msun\,pc$^{-3}$). The purple area represents the values estimated by \protect\cite{holmberg_local_2000} (this appears as a diagonal stripe since these authors only constrained $\rhodm + \rhos$). The blue-dashed lines show our priors on $\rhos$. The horizontal dashed orange line marks the Standard Halo Model value of $\rhodm$ (0.008 \msun\,pc$^{-3}$). The green area marks the range of $\rhodm^\mathrm{KG}$ we extrapolate from KG91. {\it Lower Panel:} The recovered dark matter density as a function of the visible matter surface density $\Sigma_\mathrm{s}$. The meaning of the symbols is the same as in the upper panel. The striped grey area is the range in the visible matter surface density determined by \protect\cite{flynn_2006}.}
\label{fig:chi}
\end{figure}

The results for the MA method applied to the real K dwarf data are shown in figure \ref{fig:chi}. The upper panel shows the density of the models explored by the MCMC (grey contours) in the $\rhos-\rhodm$ plane; the median with $90\%$ errors is shown by the black dot and corresponds to $\rhodm= 0.025^{+ 0.014}_{- 0.013}$\,\msun\,pc$^{-3}$ ($0.95^{+0.53}_{-0.49}$\,GeV\,cm$^{-3}$)\footnote{1\,GeV\,cm$^{-3}\simeq 0.0263158$\msun\,pc$^{-3}$.}; adding the rotation curve correction, we obtain $\rhodm=0.022^{+0.015}_{-0.013}$\msun pc$^{-3}$ ($0.85^{+0.57}_{-0.50}$\,GeV\,cm$^{-3}$, see the red dot in figure \ref{fig:chi}). 

We expect the MA method to primarily constrain the total matter density in the plane $\rhos+\rhodm$, which means that we should see an oblique degeneracy between $\rhos$ and $\rhodm$ in the $\rhos-\rhodm$ plane, as was observed for the simulation (see figure \ref{fig:KGvsMA}). However, unlike the simulation that has only one visible matter component in the disc, our real-data mass model comprises some 15 separate components with different scale heights. This introduces new freedoms that wash out the oblique degeneracy in the $\rhos-\rhodm$ plane. If we plot instead, however, the {\it total surface density} of visible matter in the disc (bottom panel of figure \ref{fig:chi}) the degeneracy is once again clearly visible as a diagonal elongation of the MCMC model density contours. Notice that many of the models explored by the MCMC lie outside of the range given by the Flynn et al. 2006 mass model (grey striped band). This simply means that the prior we placed on $\Sigma_\mathrm{s}$ is not very strong. However, the full area explored is in good agreement with the more conservative measurements of the total visible matter surface density by \cite{kg_1991} (hereafter KG91), namely $\Sigma_\mathrm{s}=48\pm 8$\,\msun\,pc$^{-2}$; and by \cite{flynn_density_1994}, $\Sigma_\mathrm{s}=49\pm 9$\,\msun\,pc$^{-2}$. Even if we include only models that lie within the grey striped region, our recovered $\rhodm$ is little affected. 

In Appendix \ref{app:low-high-z}, we test the robustness of our result, exploring how it changes if one considers either only the low $z$ bins ($0.2<z<0.7$\,kpc) or only the high $z$ bins ($0.6<z<1.2$\,kpc). We find that the low $z$ data do not provide any information about $\rhodm$, but they still favour low $\Sigma_\mathrm{s}$ compared to the prior we imposed. The low $z$ bins present a noisier and not monotonically increasing velocity dispersion, however they are better constrained and dominate the $\chi^2$ fit. Using the high $z$ data, we lose information about the disc and $\Sigma_\mathrm{s}$ settles into the centre of its prior distribution; this leads to a systematically lower $\rhodm$, since the sum of the two is well constrained. This tells us that the origin of our high $\rhodm$ is the slightly lower $\Sigma_\mathrm{s}$ required by the velocity dispersion data near the plane.

We also tested the effect of changing the assumed errors on the baryonic mass model. We first reduce them to an optimistic 10\% error for the stellar normalisations and 30\% for the gas normalisations; this has little effect on the resulting determination of $\rhodm$ because sufficient freedom remains in our mass model to allow degeneracies between $\rhodm$ and the baryonic components (this reflects the broken degeneracy seen in figure \ref{fig:chi}). We then increased the errors by removing the prior on $\rhos(0)$ altogether. This also has a small effect on our recovered $\rhodm$. This confirms the results shown in Appendix \ref{app:low-high-z} that it is the velocity dispersion data, not our prior that are constraining our mass model. The low $z$ ($\lesssim 500$\,pc) data constrain the surface density profile of the disc, while the high $z$ data ($\gtrsim 500$\,pc) are dynamically sensitive to dark matter.

The green horizontal stripe in the upper panel of figure \ref{fig:chi} marks the value of $\rhodm$ we extrapolate from KG89II and KG91. KG91, using the same K dwarf data analysed in this paper, determine the total dynamical surface density up to 1.1\,kpc: $\Sigma_\mathrm{dyn}=71\pm 6$\,\msun\,pc$^{-2}$. If we subtract from this the contribution of the observed visible matter $\Sigma_\mathrm{s}=48\pm 8$\,\msun\,pc$^{-2}$, we can calculate $\rhodm$, assumed to be constant in the range $0 < z < 1.2$\,kpc, as: 

\begin{equation} 
\rhodm^\mathrm{KG} = \frac{\Sigma_\mathrm{dyn}-\Sigma_\mathrm{s}}{2\cdot 1100}\,
\end{equation}
This gives: $\rhodm^\mathrm{KG} = 0.010\pm 0.005$\,\msun\,pc$^{-3}$. 

Our new result from our MA method is in tension with $\rhodm^\mathrm{KG}$, obtained from the same data set. This could owe either to our different distance calibration or to our new MA method that does not require any assumption about the separability of the distribution function (see Section \ref{sec:Nbody}). In figure \ref{fig:newdata}, we already showed that our new distance calibration does not significantly affect the velocity dispersion and the density fall off of the K dwarfs. In addition, in Appendix \ref{app:met}, we explore the effect of using a constant metallicity gradient for the K-dwarfs of $-0.3$\,dex\,kpc$^{-1}$, exactly as assumed by KG89II. This metallicity distribution is not compatible with modern data; we only use it to illustrate the sensitivity of our results to the MDF of the K dwarf stars, and to fully understand why our determination of $\rhodm$ is larger than that of KG91. Using the KG89II's MDF, our recovered value of $\rhodm$ is slightly smaller and therefore in better agreement with KG89II and KG91. However, our median recovered value remains significantly larger than the upper bound of the KG91 result. This suggests that our new distance determinations are not the primary reason for the systematic shift. 

\begin{figure}
\centering
\begin{tabular}{c}
\includegraphics[width=0.5\textwidth]{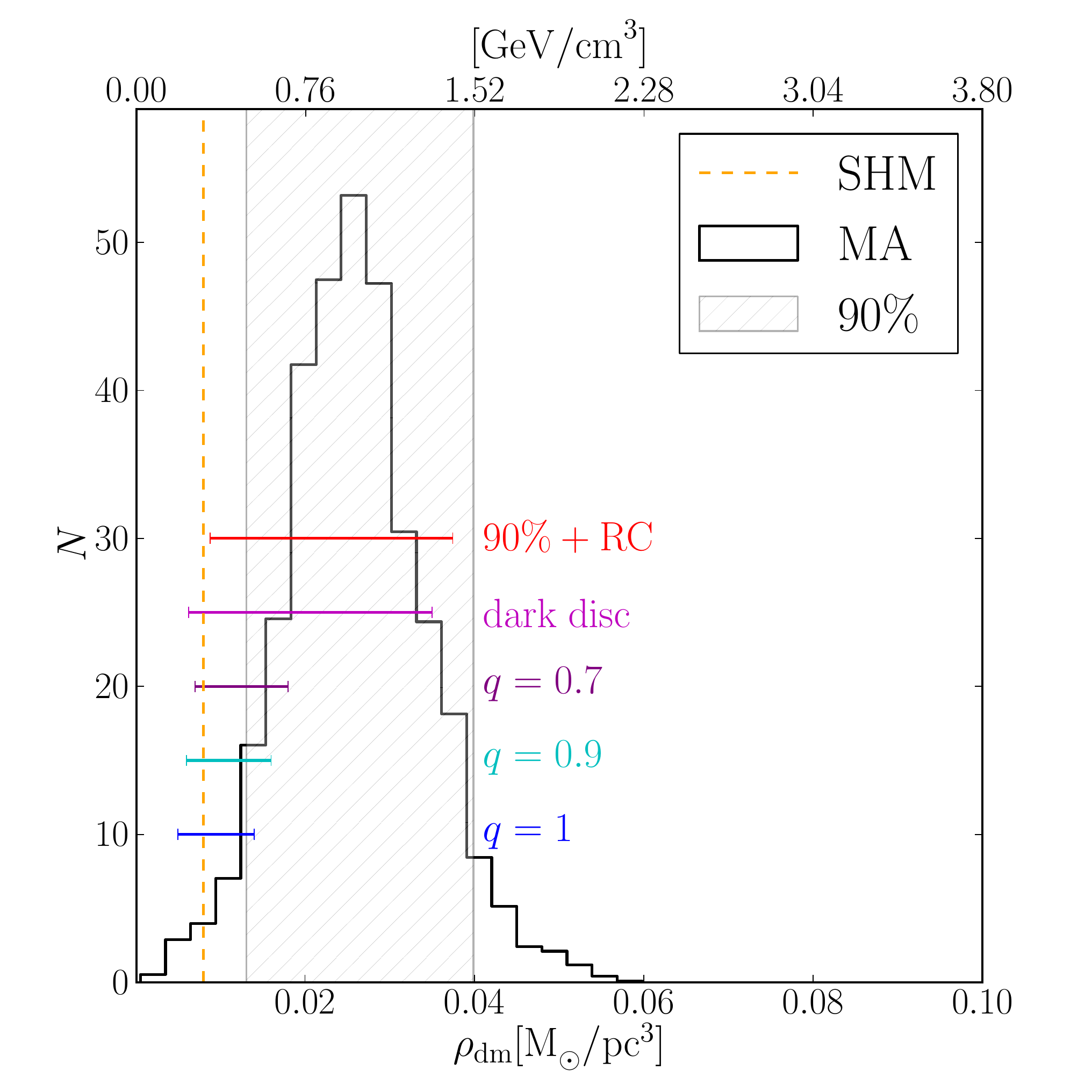}\\
\includegraphics[width=0.5\textwidth]{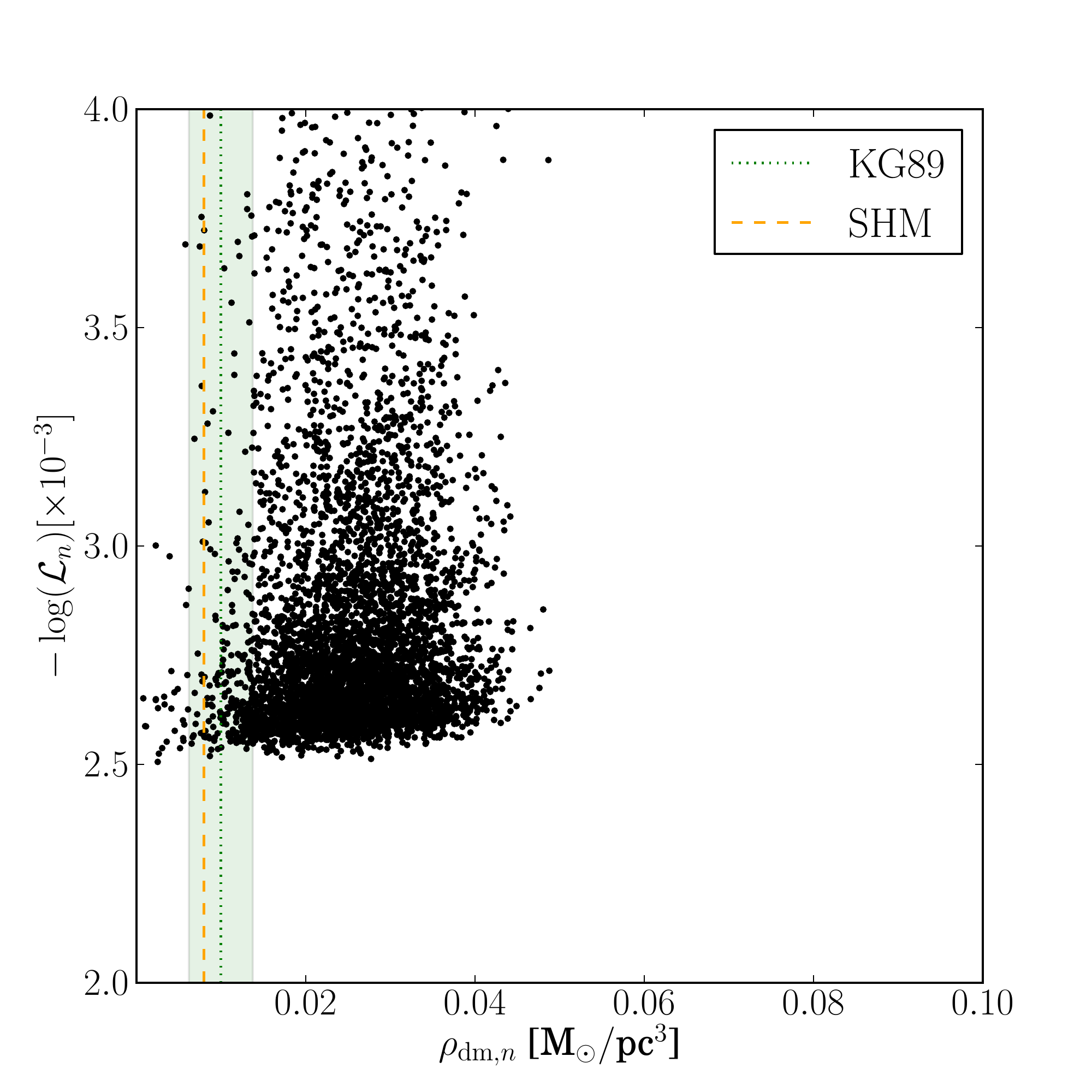}
\end{tabular}
\caption{{\it Upper panel:} A histogram of the recovered $\rhodm$ from our MCMC chains for the MA method applied to the real K dwarf data. The striped grey area is the $90\%$ confidence interval. The orange dashed line is the SHM value of $\rhodm$; the blue, cyan and purple error bars correspond to the value of $\rhodm$ obtained by \protect\cite{iocco_2011} from a combination of microlensing and rotation curve data, with a spherical halo (potential flattening $q=1$) and two oblate halos, with $q=0.9$ and $q=0.7$, respectively. The magenta error bar is the value of $\rhodm$ expected if the Milky Way has a dark disc contributing $0.25-1.5$ times the density of the (spherical) halo. The red error bar corresponds to our result after adding the rotation curve correction. {\it Lower panel:} The effect of assuming a separable distribution function. Each dot shows the likelihood of a given MCMC model calculated assuming a separable distribution function. Notice that the assumption of separability biases the result towards low $\rhodm$. The orange and the green dashed lines have the same meaning as in the upper panel of \ref{fig:chi}.}
\label{fig:KGlik}
\end{figure}

In our tests on the N-body simulation (Section \ref{sec:Nbody}), we showed that, when the distribution function of the tracers is not separable, the method adopted by KG89II leads to a systematic underestimate of $\rhodm$. In the lower panel of figure \ref{fig:KGlik}, we plot the likelihood $\mathcal{L}_n$ of each model explored by our MCMC -- calculated through equation \ref{eqn:LIK} -- against the corresponding value of $\rhodm$. Unlike the similar plot for our simulation data (figure \ref{fig:KGvsMA}, bottom panels), there is now a significant vertical dispersion in the models. This owes to the increased freedom present in our 15-parameter mass model for the real-data. However, the highest likelihood models (the bottom envelope of points in the plot) show a similar trend as seen for the simulation data: higher likelihood models have systematically smaller $\rhodm$. We conclude that the primary difference for our larger value of $\rhodm$ as compared to KG89II is that our MA method requires no assumption about the separability of the distribution function. 

In the upper panel of figure \ref{fig:KGlik}, we plot a histogram of $\rhodm$ from all the models explored by the MCMC. The striped area is the $90\%$ confidence interval (corresponding to the black dot of figure \ref{fig:chi}); the result including the rotation curve correction is shown by the red error bar. Notice that our 90\% {\it lower bound} is larger than the Standard Halo Model\footnote{The SHM is an isothermal sphere model for the Milky Way's dark matter halo with a value of the dark matter velocity dispersion assumed to be $\sigma_\mathrm{iso}\simeq 270$\,km\,s$^{-1}$.} typically assumed in the literature (marked by the vertical dashed orange line). For a comparison, we plot the ranges of $\rhodm$ at the solar radius obtained by \cite{iocco_2011}, combining microlensing and rotation curve measurements, and using different halo models: the blue error bar corresponds to a spherical halo, while the cyan and purple bars correspond to oblates halos with potential flattening $q=0.9$ and $q=0.7$, respectively. The magenta bar represents the dark matter density in presence of a dark disc, contributing $0.25-1.5$ times the dark matter (spherical) halo density, as predicted by \cite{read_dark_2009}. 

From figure \ref{fig:KGlik}, we can see that our recovered density is in mild tension with the result for a spherical Milky Way halo. Moving to an oblate halo significantly reduces this tension, however a flattening of $q = 0.7$ is likely inconsistent with measurements of the halo shape from the Sagittarius stream of stars  \citep[e.g.][]{Ibata_2001}. If we wish to explain our median value for $\rhodm$ that is very much larger than the canonical SHM value assumed in the literature, we require a local disc of dark matter that raises $\rhodm$ without significantly altering the rotation curve. Interestingly, our median value is in excellent agreement with the range of dark discs predicted for our Galaxy by \cite{read_thin_2008,read_dark_2009}.

\section{Discussion and conclusions}\label{sec:discuss+concl}
We have presented a new measurement of the local matter and dark matter densities from the kinematics of K dwarf stars near the Sun. We presented a new photometric distance calibration for the the K dwarf data of KG89II (the KG data), derived using modern survey catalogues and the Hipparcos satellite data. We then used these data as tracers of the local gravitational potential to calculate the visible ($\rhos$) and dark matter ($\rhodm$) densities at the solar position $R_\sun$ and the surface density of the Milky Way disc up to 1.1\,kpc above the plane ($\Sigma_s$). 

To determine $\rhodm$ and $\rhos$, we applied our new mass modelling method (presented already in Paper I) that relies on a minimum set of assumptions (the MA method) to the rejuvenated KG data. The key advantages of our new method are that: (i) we do not require any hypothesis about the shape of the tracers' velocity distribution function; (ii) we use a MCMC to marginalise over uncertainties in the distances and velocities of the tracer stars, and the underlying baryonic mass model for the visible disc; and (iii) we require no prior from the Milky Way rotation curve as has been commonly used in previous works. This latter means that we can compare our determination to that derived from the rotation curve to constrain the Milky Way halo shape. We used a dynamically evolved high resolution N-body simulation of a Milky Way-like galaxy as a mock data set to test our MA method, finding that we could correctly recover $\rhodm$ and $\rhos$ within our 90\% confidence interval (for eight sample Solar neighbourhood-like volumes) even in the face of disc inhomogeneities, non-isothermal tracers, asymmetric distance errors and a non-separable tracer distribution function. Furthermore, we confirmed the result from our Paper I that assuming a separable distribution function (as has been typically done in the modern literature) leads to a biased determination of $\rhodm$.

Applying our MA method to the K dwarf data, we obtain a new measurement of the local dark matter density: $\rhodm= 0.025^{+ 0.014}_{- 0.013}$\,\msun\,pc$^{-3}$ ($0.95^{+0.53}_{-0.49}$\,GeV\,cm$^{-3}$); which, adding a correction for the local non-flatness of the rotation curve correction term ($\simeq -0.0033\pm0.0050$, see Sections \ref{sec:MA} and \ref{sec:Res}), gives: $\rhodm=0.022^{+0.015}_{-0.013}$\msun pc$^{-3}$  ($0.85^{+0.57}_{-0.50}$\,GeV\,cm$^{-3}$). Our new value is systematically larger than the results from KG89II and KG91 derived from the same data. We show that this primarily owes to our new MA modelling method (and the fact that it does not assume a separable distribution function for the tracers); our new distance determination for the K dwarfs plays a more minor role. At the same time we determine a value of the local visible matter density of $\rhos = 0.098^{+0.006}_{-0.014}$\,\msun\,pc$^{-3}$ that largely reflects the prior from our baryonic mass model. 

Our error bars are larger than is often quoted in the literature, however they reflect the full combination of model systematic, measurement and statistical uncertainties. Other recent determinations either rely on the rotation curve and therefore a strong assumption about the Milky Way halo shape, or require a large number of assumptions with associated (and typically unmodelled) systematic errors.

In addition to measuring $\rhodm$ and $\rhos$, we also obtain an estimate of the baryonic disc mass up to $z = 1.1$\,kpc above the disc plane: $\Sigma_\mathrm{s}=45.5^{+5.6}_{-5.9}$\msun pc$^{-2}$ at 90\% confidence. This is slightly lower than the mean of the prior from our baryonic mass model: $\Sigma_\mathrm{vis} = 49.4 \pm 4.6$\,\msun\,pc$^{-2}$. Splitting the number into the contribution from stars and stellar remnants: $\Sigma_* = 33.4^{+5.5}_{-5.2}$\,\msun\,pc$^{-2}$ and gas: $\Sigma_\mathrm{g} = 12.00^{+1.9}_{-2.0}$\,\msun\,pc$^{-2}$, we see that our model favours slightly lower surface density in both the gas and the stars than the mean of our priors ($\Sigma_{\mathrm{g}}^\mathrm{obs}=13.3\pm3.4$\,\msun\,pc$^{-2}$, $\Sigma_{*}^\mathrm{obs}=36.1\pm3.0$\,\msun\,pc$^{-2}$). 

It is this tendency for our models to favour lower disc surface density that leads to our high median value for $\rhodm$ (see figure \ref{fig:chi} and Appendix \ref{app:low-high-z}). Unfortunately, current estimates of the stellar and gaseous inventory in the Solar neighbourhood are too uncertain to confirm or rule out our favoured $\Sigma_\mathrm{s}$ \citep[e.g.][]{bovy_rix_2012,ferriere2001}.

Our median value of the local dark matter density is larger at 90\% confidence than the Standard Halo Model value of 
$\rhodm^\mathrm{SHM}=0.008$\,\msun\,pc$^{-3}$ ($0.30$\,GeV\,cm$^{-3}$) usually adopted in the literature. This could be a statistical fluctuation; in one out of eight patches in our simulated mock data, our method overestimated $\rhodm$ by $\sim90\%$. However, if our high median value is confirmed by future data then it has some interesting implications. Firstly, it is particularly important for direct detection experiments because it implies a larger flux of dark matter particles and therefore a greater chance of detection. Secondly, our result is at mild tension with the value of $\rhodm^\mathrm{ext}$ extrapolated from the rotation curve measurements, assuming a spherical dark matter halo. This suggests that the halo of our Galaxy is oblate and/or that we have a disc of dark matter, as predicted by recent cosmological simulations (see upper panel of figure \ref{fig:KGlik}).

\section*{Acknowledgments}
We would like to thank Lan Zhang for kindly supplying the SDSS data we used to calculate the K dwarfsÕ metallicity distribution function. We thank Glenn Van de Ven for providing us the biweight code for the calculation of the velocity dispersions and for useful discussions. We would like to thank the referee Chris Flynn for a careful reading of the manuscript and useful comments. Finally, we would like to thank Fabio Iocco, Miguel Pato and Jo Bovy for useful discussions. We thank Scott Tremaine for very useful comments that aided the clarity of our work.
Justin I. Read would like to acknowledge support from SNF grant PP00P2\_128540/1.

\bibliographystyle{mn2e}
\bibliography{paper2bibl}

\appendix
\section{Testing the robustness of the MDF}\label{app:met}
\begin{figure*}
\centering
\begin{tabular}{cc}
\includegraphics[width=0.45\textwidth]{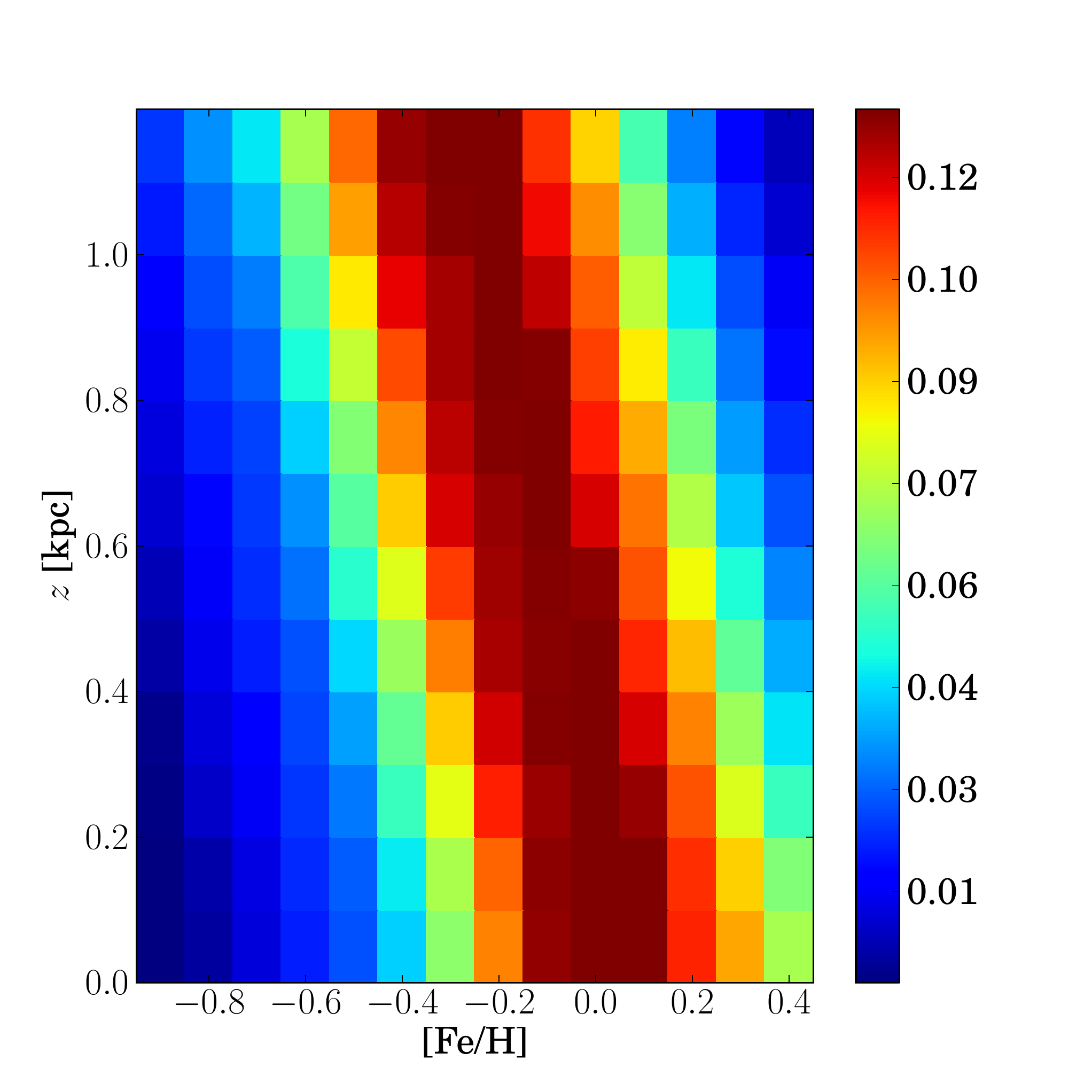} &
\includegraphics[width=0.45\textwidth]{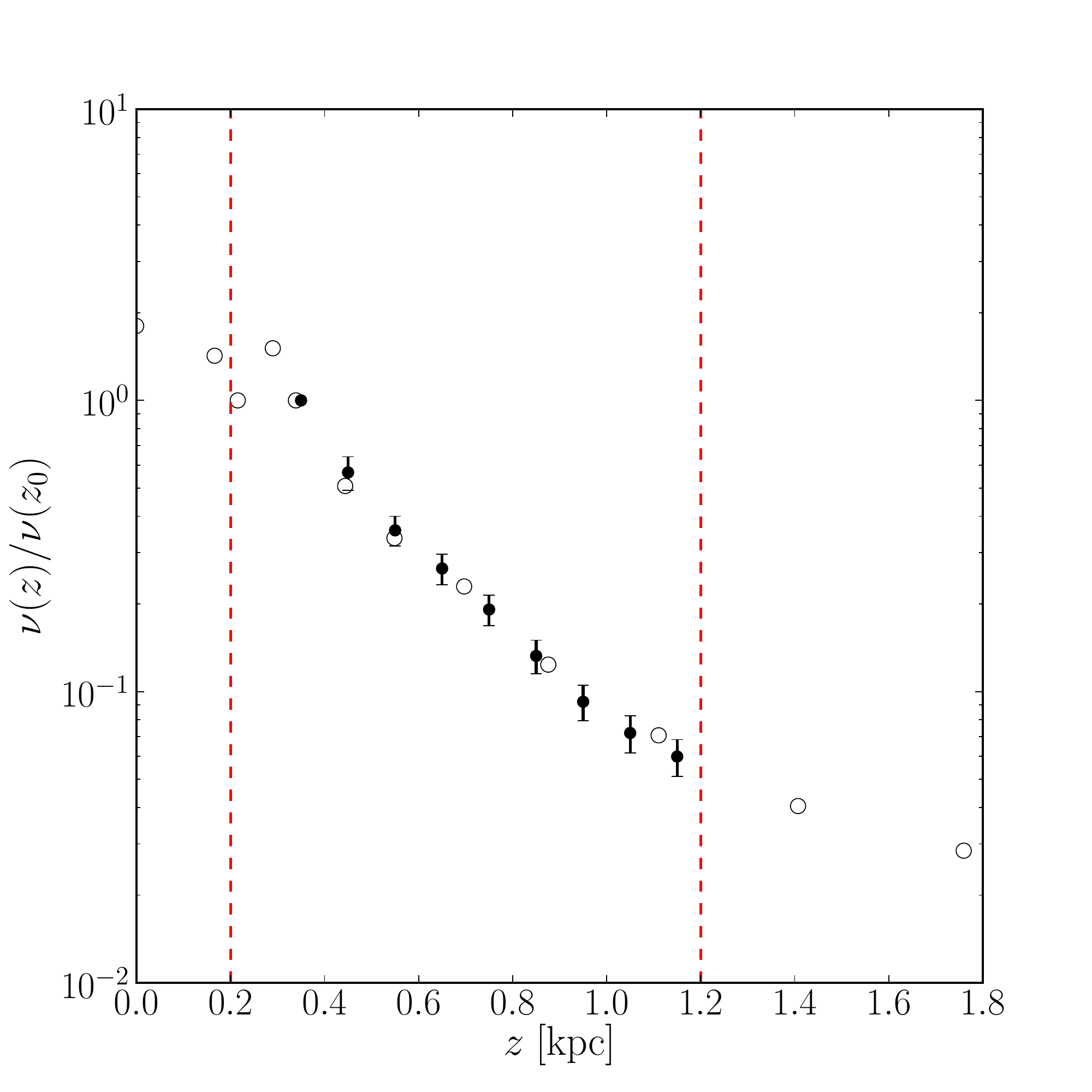} \\
\includegraphics[width=0.45\textwidth]{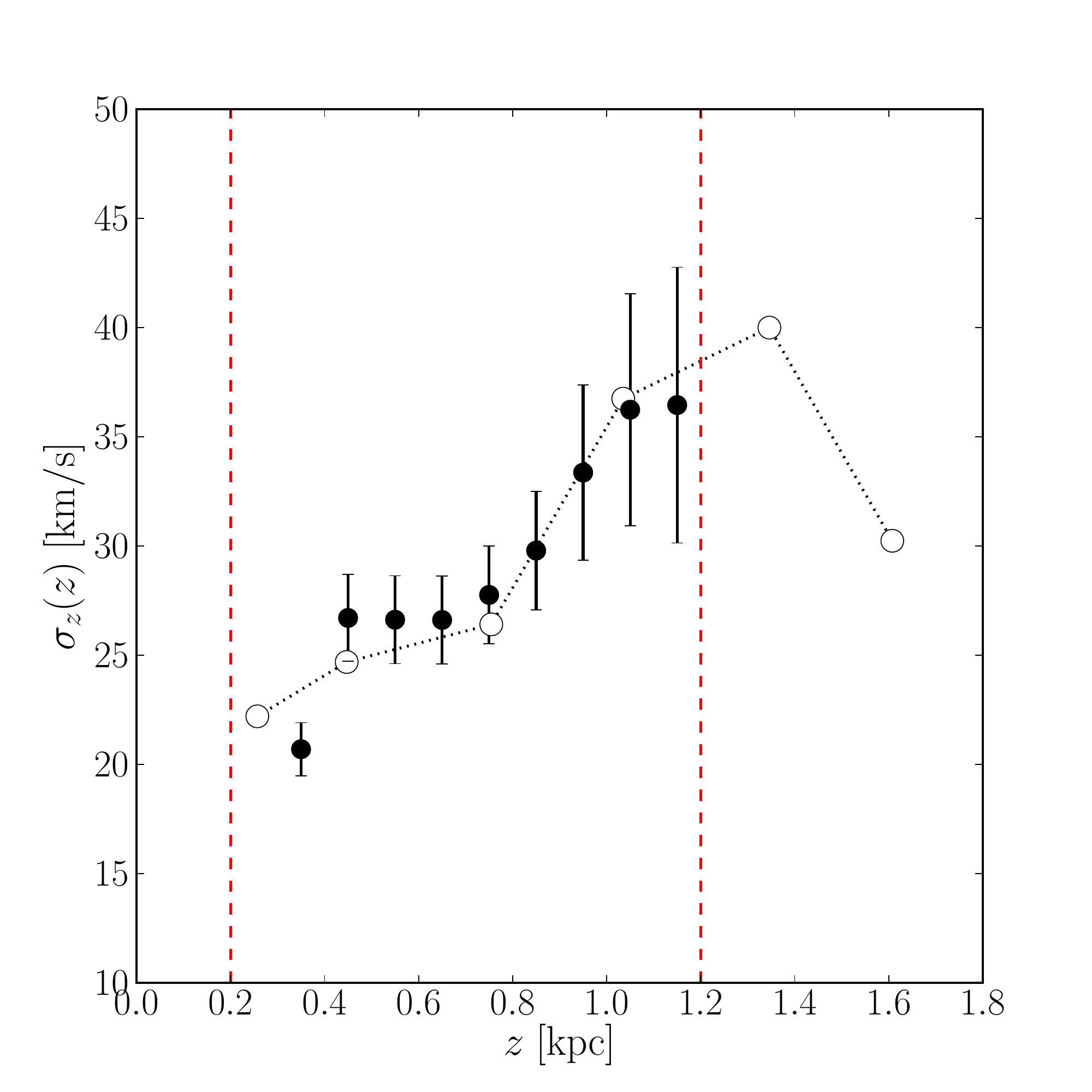} &
\includegraphics[width=0.45\textwidth]{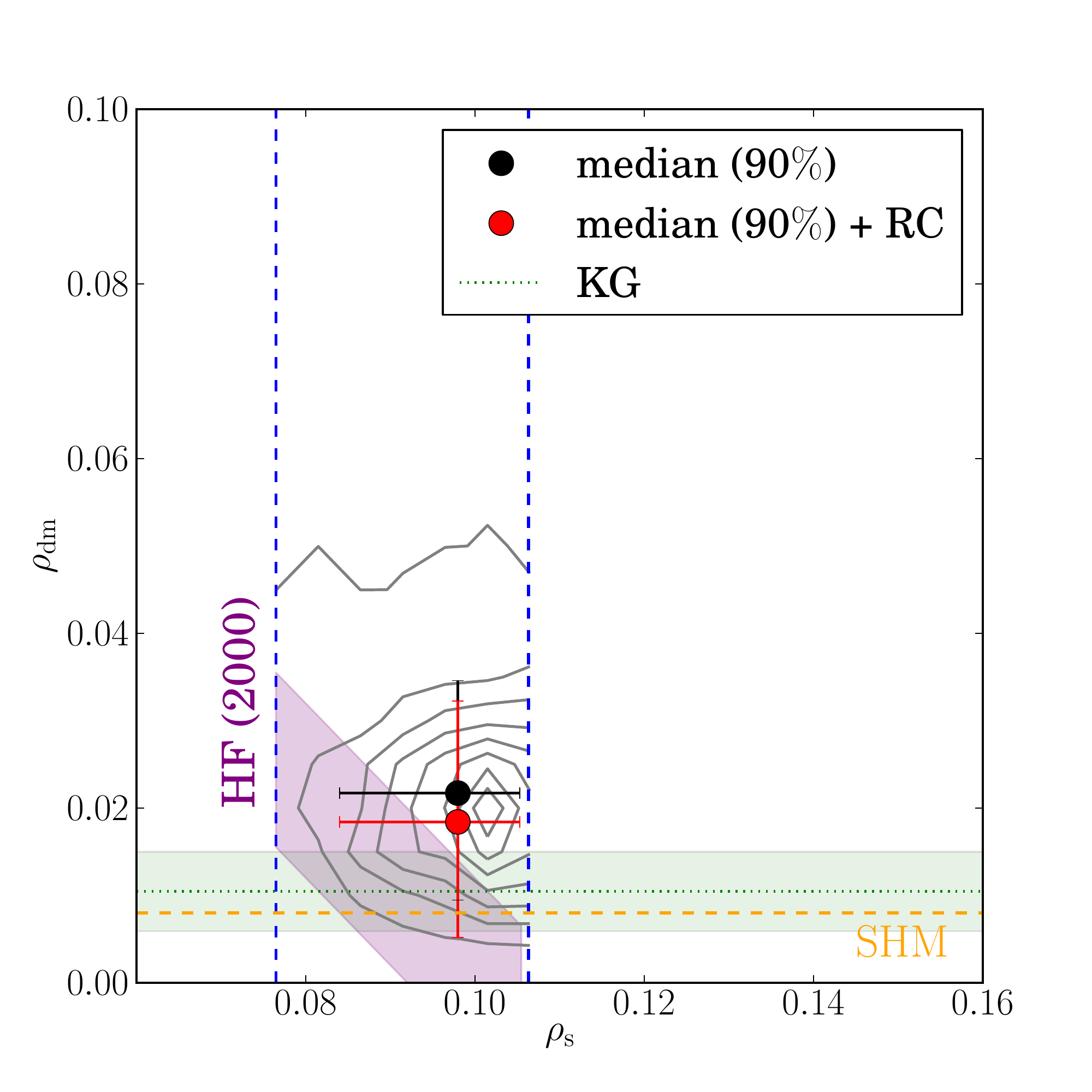}
\end{tabular}
\caption{{\it Left upper panel:} The MDF adopted by KG89II: a constant metallicity gradient with height of 0.3\,dex\,kpc$^{-1}$ normalised to a metallicity of 0.\,dex at $z=0$\,kpc. {\it Right upper panel}: Stellar density profile derived from Monte Carlo sampling the probability distribution of $z$ (dots with error bars). As a comparison, the density profile from KG89II is plotted as empty circles. The two red dashed lines show the completeness range. {\it Lower left panel}: The vertical velocity dispersion as a function of $z$. The black dots with error bar are derived from the probability distribution of $z$. {\it Lower right panel}: The projection of MCMC models onto the $\rhodm$-$\rhos$ plane for our MA method applied to the K dwarf data, using the MDF shown in the top left panel to determine the distances. The colours and the symbols are as in figure \ref{fig:chi}.}
\label{fig:KGmet}
\end{figure*}

The most uncertain quantity in our re-analysis of KG89II's data is the variation of the K dwarfs' metallicity distribution function with $z$, $Q(\mathrm{[Fe/H]}(z),z)$. In this appendix, we investigate how the adopted metallicity distribution function affects the result of our analysis by exploring a different model for this function. We use the gradient adopted by KG89II, i.e. $-0.3$\,dex\,kpc$^{-1}$, and set the mean metallicity to 0 at $z=0$ (see upper left panel of figure \ref{fig:KGmet}). This MDF is {\it not} compatible with modern metallicity data for the K dwarfs (see figure \ref{fig:methist}); we use it simply to illustrate our sensitivity to the assumed MDF, and to aid comparisons with the earlier KG89II results. 

In figure \ref{fig:KGmet}, we show the velocity dispersion (upper right panel) and the density fall off (lower left panel) of the ttracers derived using the above MDF. In the lower right panel of figure \ref{fig:KGmet}, the recovered dark and visible matter density are shown. The value of $\rhodm$ obtained is slightly lower than that derived using our default MDF, namely $\rhodm=0.022^{+0.013}_{-0.012}$\,\msun\,pc$^{-3}$, or $\rhodm=0.018^{+0.014}_{-0.013}$\,\msun\,pc$^{-3}$ including the rotation curve correction. However, our median value for $\rhodm$ is still high and the overall result remains in tension with the SHM value (even when using this incorrect MDF). We conclude that our results are robust to plausible variations in the assumed MDF.

In addition to the above, we also tested the impact of a different extinction value used in equation \ref{eqn:d} on our results. We chose $A_V=0$ and $A_V=0.1$. For $A_V=0$ ($A_V=0.1$) the distances are slightly overestimated (underestimated) with respect to the extinction values considered in this article. This does not affect much the density fall off, but it translates mainly in a slightly flatter (steeper) velocity dispersion. This leads to a small decrease (increase) of the recovered value of $\rhodm$. However the impact of the extinction is very small compared to the previous test shown in this Appendix.

Finally we tested the impact on our distance calibration of the larger scatter in equation \ref{eq:poly} obtained using the more modern K dwarf catalogue from \cite{casagrande2007} instead of \cite{kotoneva_2002}. The two studies are very much compatible, but the scatter in equation \ref{eq:poly} of 0.03\,mag increases to 0.27\,mag when the newer data are considered. We used the 104 K dwarf stars from \cite{casagrande2007} to build the relationship between $M_V$ and ($B-V$ , [Fe/H]), similar to equation \ref{eq:poly}, but using the polynomial down to $a_{2,1} (B-V)^2$[Fe/H]. Unlike \cite{kotoneva_2002}, the new catalogue also provides the error on the parallax, therefore we use a $\chi^2$ instead of a simple linear-least-square fit. To account for the 0.27\,mag uncertainties on $M_V$, we convolved the distance probability distribution function with a Gaussian of width $\sigma = 13$\% (corresponding to 0.27\,mag in $M_V$). The resulting density and velocity dispersion profiles are unchanged in the range of $z$ of interest, with only a negligible increase in the uncertainties. Unlike in the case of the extinction test, where distances were (slightly) systematically overestimated or underestimated, the magnitude scatter has only the effect of slightly increasing the random errors. The effect is weak so as long as a sufficiently large number of stars is used per bin, and the scatter is not too large. For this reason our determination of the local dark matter density is not affected by the increased scatter found by \cite{casagrande2007}.

\newpage
\section{Exploring the robustness of our $\rhodm$ determination}\label{app:low-high-z}
\begin{figure*}
\centering
\begin{tabular}{ccc}
\large{$0.2<z<0.7$\,kpc} & \large{$0.2<z<1.2$\,kpc} &\large{$0.6<z<1.2$\,kpc} \\
\includegraphics[width=0.33\textwidth]{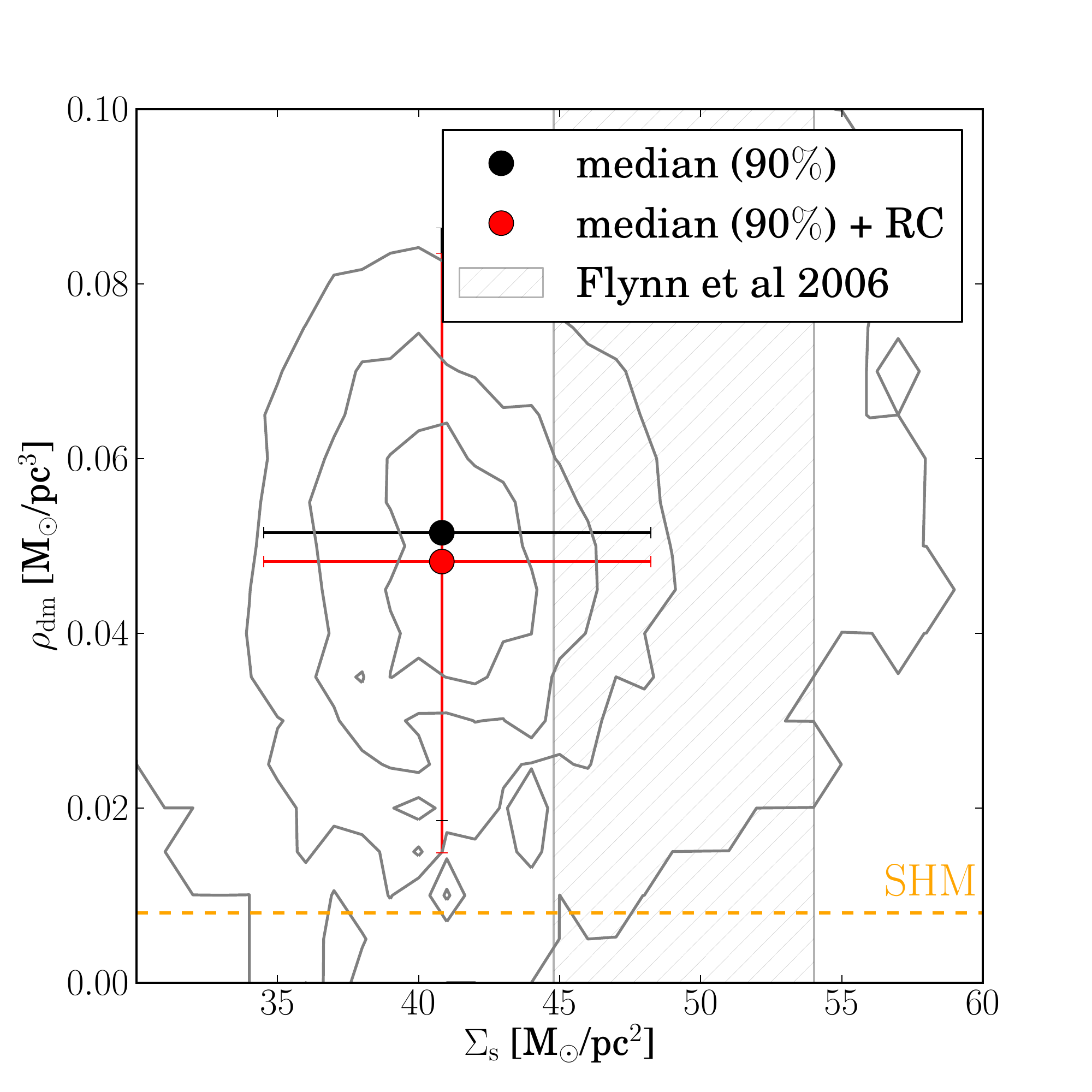} &
\includegraphics[width=0.33\textwidth]{bmet_Sigcont.pdf} &
\includegraphics[width=0.33\textwidth]{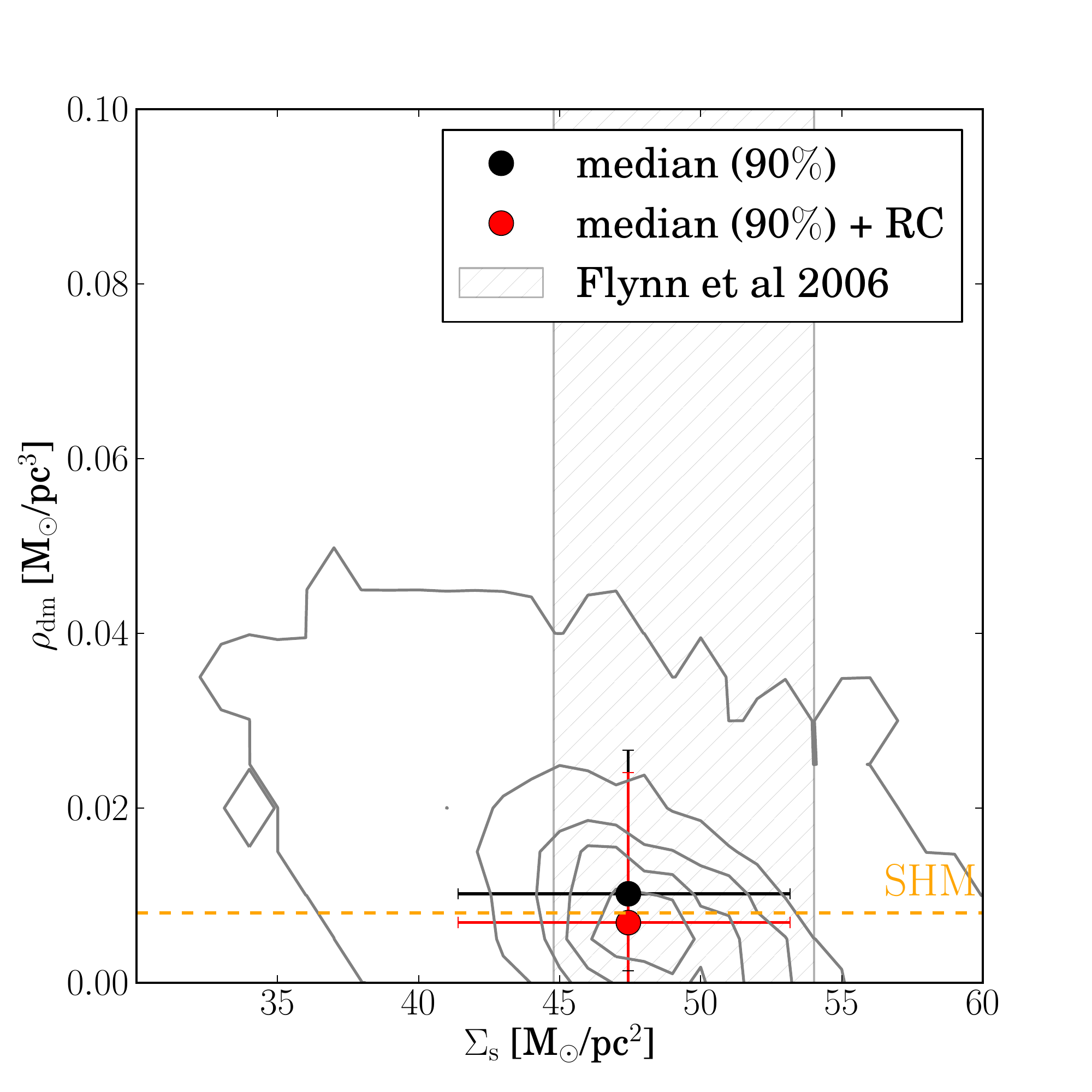}
\end{tabular}
\caption{The recovered dark matter density as a function of the visible matter surface density $\Sigma_\mathrm{s}$. {\it Left panel:} considering only low $z$ bins ($0.2<z<0.7$\,kpc); {\it central panel:}  considering the full $z$ range: $0.2<z<1.2$\,kpc (as lower panel of figure \ref{fig:chi}); {\it right panel:} considering only high $z$ bins ($0.6<z<1.2$\,kpc). The meaning of the symbols is the same as in figure \ref{fig:chi}. The striped grey area is the range in the visible matter surface density determined by \protect\cite{flynn_2006}, that we used as a (weak) prior.}
\label{fig:low-high-z}
\end{figure*}

In this Appendix, we explore the robustness of our determination of $\rhodm$ by analysing a low $z$ ($0.2 < z < 0.7$\,kpc) and a high $z$ ($0.6 < z < 1.2$\,kpc) subset of the KG data. The results are shown in Figure \ref{fig:low-high-z}. If we consider only the low $z$ data (left panel), there is no information about the dark matter density. However, the data still favour low $\Sigma_\mathrm{s}$ compared to our prior. If only the high $z$ data are used (right panel), we lose the information about the disc and $\Sigma_\mathrm{s}$ settles into more or less the centre of its prior distribution. This leads to a systematically lower $\rhodm$. The above suggests that the origin of our high median $\rhodm$ is the lower $\Sigma_\mathrm{s}$ favoured by the velocity dispersion data close to the plane.

\label{lastpage}
\end{document}